\documentclass[%
 aip,
 amsmath,amssymb,
 reprint,%
 groupedaddress,
]{revtex4-1}

\usepackage{graphicx}
\usepackage{dcolumn}
\usepackage{bm}

\usepackage[utf8]{inputenc}
\usepackage[T1]{fontenc}
\usepackage{mathptmx}
\usepackage{etoolbox}
\usepackage{amsmath}
\usepackage{mathtools}
\usepackage{cases}
\usepackage{comment}
\newtagform{nowidth}{\llap\bgroup(}{)\egroup}
\usepackage{soul}
\usepackage{xcolor}

\usepackage{color}
\newcommand{\comm}[1]{{{\textcolor{red}{\bf #1}}}} 

\DeclareMathAlphabet{\mathdutchcal}{U}{dutchcal}{m}{n}
\SetMathAlphabet{\mathdutchcal}{bold}{U}{dutchcal}{b}{n}
\DeclareMathAlphabet{\mathdutchbcal}{U}{dutchcal}{b}{n}

\makeatletter
\def\@email#1#2{%
 \endgroup
 \patchcmd{\titleblock@produce}
  {\frontmatter@RRAPformat}
  {\frontmatter@RRAPformat{\produce@RRAP{*#1\href{mailto:#2}{#2}}}\frontmatter@RRAPformat}
  {}{}
}%
\makeatother

\begin{document}

\preprint{AIP/123-QED}

\title[Phase space dynamics of unmagnetized plasmas: collisionless and collisional regimes]{Phase space dynamics of unmagnetized plasmas: collisionless and collisional regimes}
\author{G. Celebre}
\author{S. Servidio}
\author{F. Valentini}
\affiliation{Dipartimento di Fisica, Università della Calabria, I-87036 Rende (CS), Italy}

\begin{abstract}
The following article has been accepted by Physics of Plasmas. After it is published, it will be found at
https://publishing.aip.org/resources/librarians/products/journals/ . Copyright (2023) G. Celebre, S. Servidio, and F. Valentini. This article is distributed under a Creative Commons Attribution (CC BY) License. \\[10 pt]
Eulerian electrostatic kinetic simulations of unmagnetized plasmas (kinetic electrons and motionless protons) with high-frequency equilibrium perturbations have been employed to investigate the phase space free energy transfer across spatial and velocity scales, associated with the resonant interaction of electrons with the self-induced electric field. Numerical runs cover a wide range of collisionless and weakly collisional plasma regimes. An analysis technique based on the Fourier-Hermite transform of the particle distribution function allows to point out how kinetic processes trigger the free energy cascade, which is instead inhibited at finer scales when collisions are turned on. Numerical results are presented and discussed for the cases of linear wave Landau damping, nonlinear electron trapping, bump-on-tail and two-stream instabilities. A more realistic situation of turbulent Langmuir fluctuations is also discussed in detail. Fourier-Hermite transform shows a free energy spread, highly conditioned by collisions, which involves velocity scales more quickly than the spatial scales, even when nonlinear effects are dominant. This results in anisotropic spectra whose slopes are compatible with theoretical expectations. Finally, an exact conservation law has been derived, which describes the time evolution of the free energy of the system, taking into account the collisional dissipation. 
\end{abstract}

\maketitle

\section{\label{sec:level1}Introduction} \label{secI}


The plasma dynamics is extraordinarily complex because particles and electromagnetic fields interact in a fully nonlinear way. This unpredictable behavior can evolve into turbulence, where energy is distributed on a wide range of scales and frequencies. As in ordinary fluids, turbulence manifests in a cascade-like process where the energy available within the system tends to transit from large to small spatial scales, until the collisions between particles dissipate available energy by increasing the temperature of the plasma.

Space plasma observations, for example, suggest that the electric and magnetic field fluctuations are highly turbulent, leading to complex interactions between fields and particles \cite{MMS}. In recent observations \cite{Klein}, a measurement of collisionless damping in heliospheric plasmas suggested an energy transfer between fields and particles, where various plasma processes and instabilities might be at work.

The plasma dynamics can be described via a kinetic model, by using the Vlasov-Maxwell system of equations \cite{VlaMax}, which determines the time evolution of the particle distribution function. An effective, quantitative analysis of the distribution function is crucial to investigate the nature of phenomena that can be observed in plasmas, such as turbulence and wave-particle interactions.

Our work is focused on the Fourier-Hermite transform (FHT) technique, a spectral analysis which proved to be particularly useful to examine the features of particle distribution function in several contexts. It has already been employed with success in previous works, as a spectral resolution method of the Vlasov-Maxwell system but also to analyze plasma simulations and spacecraft data (see Refs.~\onlinecite{MMS,FH1,FH2,FH3,FH4,FH5}). In particular, we apply this technique to several simulated plasma states which can be described by Vlasov-Maxwell equations. The main novelty of the present work is that we employ this spectral analysis to a very large class of regimes, in order to highlight similarities and differences amongst plasma states through the Fourier-Hermite spectra.

We investigate plasmas composed of protons and electrons that are unmagnetized and overall neutral, with high-frequency perturbations that leave protons fixed with constant particle density $n_0$. In this configuration, electrons can move only along one direction, so the variables are the spatial coordinate $x$, the velocity component $v$ parallel to $x$, and time $t$. Under these hypotheses, Vlasov-Maxwell equations turn into the 1D-1V (one dimension in physical space and one dimension in velocity space) Vlasov-Poisson (VP) system \cite{VlaPoi,EulColl,NumRec}, which, in presence of electron-electron collisions, can be written as: 
\begin{gather}
\label{sist1Dcolla}
\frac{\partial f}{\partial t} + v \frac{\partial f}{\partial x} - \frac{e}{m_e} E \frac{\partial f}{\partial v} = \frac{\partial f}{\partial t} \bigr\rvert_{coll}, \\[5pt]
\label{sist1Dcollb}
\frac{\partial E}{\partial x} = 4 \pi e \left( n_0 - \int_{-\infty}^{+\infty} f dv \right),
\end{gather}
where $f(x,v,t)$ is the electron distribution function, $E(x,t)$ is the electric field, while $e$ and $m_e$ denote the elementary charge and the electron mass, respectively. The collisional term is represented by the right-hand side of Eq.~(\ref{sist1Dcolla}), modeled with the Dougherty operator \cite{Dough1,Dough2}. Generally, it is possible to find electrons in an equilibrium state of the VP system such that $E = 0$ and $f = f_{eq} (v)$. This allows us to investigate the relationship between an initial equilibrium perturbation and the subsequent time evolution of the electron distribution and electric field.

We examine many regimes described by Eqs.~(\ref{sist1Dcolla})-(\ref{sist1Dcollb}) through direct numerical simulations, by varying the initial conditions, by comparing the collisionless ($\partial f / \partial t \rvert_{coll} = 0$) with the collisional ($\partial f / \partial t \rvert_{coll} \neq 0$) cases, and by going from linear to fully turbulent states. In the majority of cases, $f_{eq}$ corresponds to the Maxwell-Boltzmann (MB) distribution, $f_{MB} (v) = n_0 \exp [-v^2 / (2 v_{th , e}^2)] / (\sqrt{2 \pi} v_{th , e})$, where $v_{th , e}$ represents the electron thermal velocity. However, other equilibrium states are considered too. In particular, by defining the equilibrium perturbation as $\delta f (x,v,t) = f(x,v,t) - f_{eq}(v)$, we explore four large classes:
\begin{itemize}
\item[(I)] a linear regime with $\delta f (t = 0) \ll f_{MB}$, by using a single-mode perturbation in space;
\item[(II)] the same as (I), but with $\delta f (t = 0) \sim f_{MB}$ (nonlinear regime);
\item[(III)] a starting small perturbation of $f_{eq} \neq f_{MB}$, that excite either the bump-on-tail \cite{bump} or the two-stream \cite{beam} instability;
\item[(IV)] a fully turbulent state, with $\delta f (t = 0) \sim f_{MB}$, by using a superposition of several Fourier modes.
\end{itemize}

One might expect that these regimes will evolve in very different ways. For example, when the effect of nonlinear terms in the Vlasov equation is sufficiently small (i.e. the typical evolution time is much shorter than electron trapping time), the linear Landau regime should adequately describe plasma behavior \cite{Landau}. Instead, when the time scales considered are comparable to the trapping time, plasma enters the nonlinear O'Neil regime \cite{ONeil}. Furthermore, the inclusion of the collisional term $\partial f / \partial t \rvert_{coll}$ in the Vlasov equation may have a significant impact on the dynamics, as recently investigated in the context of space plasma turbulence via full Boltzmann-Maxwell simulations \cite{FouHer}.

In this context, a thorough comprehension of the phase space dynamics of electrostatic plasmas requires an appropriate spectral decomposition, for both the physical and velocity subspaces, like the FHT algorithm. Starting from the discrete electron distribution function, we apply the Hermite transform (HT) in velocity space by projecting $f$ on the Gauss-Hermite orthonormal functions $\psi_m$'s, obtaining their relative weights $f_m (x,t)$; these are then processed by the fast Fourier transform (FFT) in physical space: the final output consists of Fourier-Hermite coefficients of the form $f_{n \; m} (t)$.

The application of HT is highly advantageous because Hermite modes $m$'s can effectively represent the enstrophy associated with distribution function perturbations (see also Ref.~\onlinecite{MMS}), which is related to the free energy of the system \cite{Budget}. The 0-th Gauss-Hermite function, for example, corresponds to the normal distribution and so indicates how much the system is close to the thermodynamic equilibrium. Otherwise, when a perturbation is applied to $f$, enstrophy (or, equivalently, free energy) propagates to modes with $m > 0$: fluctuations with finer velocity scales excite modes with higher $m$. This property of the transform enables effective visualization of enstrophy cascades in velocity space \cite{MixAdv, MMS}. Note also that there is a correlation between $f_m$ and the $m$-th moment of the distribution, so it is possible to analyze the first Hermite coefficients to get information about density, temperature, heat flux and so on.

The completion of the spectral analysis through FFT permits to analyze the classical physical space cascade, establishing important analogies with fluids. The combined application of these two transforms allows for a quantitative comparison between the cascade in the velocity space and the cascade in the physical space. In particular, it can visualize, for each regime, which cascade reaches more quickly the smallest scales available and which one is suppressed more effectively by the collisional term. The synergy between HT and FFT might finally allow a statistical description of the full phase space, to understand whether a self-similar, universal scaling might occur in collisionless plasmas \cite{MixAdv, inter1, inter2}.

A further study of the dynamics of the simulated regimes is performed by calculating the terms of a free energy conservation law at different times during the numerical runs. This law can be expanded in series to get an approximated budget equation which estimates the contribution of equilibrium perturbations in the total free energy of the system. A confrontation between the exact and the approximated law permits quantifying the error due to the series truncation.

The paper is organized as follows. Sec.~\ref{secII} is dedicated to a description of the numerical code that solves the VP system. Here we also describe the procedure of the FHT. For each simulation, we provide the details of the initial conditions. The main results of our simulation campaign are reported in Sec.~\ref{secIII}. In Sec.~\ref{secIV} we discuss our conclusions.

\section{The numerical algorithm} \label{secII}
We solve numerically the 1D-1V collisional Vlasov-Poisson system of equations in Eqs.~(\ref{sist1Dcolla})-(\ref{sist1Dcollb}) for the electron distribution function $f(x,v,t)$ and the electric field $E(x,t)$. These equations can be rewritten in dimensionless units as follows:
\begin{gather} 
\label{adim1}
\frac{\partial f}{\partial t} + v \frac{\partial f}{\partial x} - E \frac{\partial f}{\partial v} = \frac{\partial f}{\partial t} \bigr\rvert_{coll}, \\[5pt]
\label{adim2}
\frac{\partial E}{\partial x} = 1 - \int_{-\infty}^{+\infty} f dv.
\end{gather}
In the above equations, velocities are scaled by the electron thermal speed $v_{th , e} = \sqrt{k_B T_{e \; 0} / m_e}$ ($T_{e \; 0}$ being the electron temperature at equilibrium and $m_e$ and $e$ the electron mass and charge respectively), lengths by the Debye length $\lambda_{D , e} = \sqrt{k_B T_{e \; 0} / (4 \pi n_0 e^2)}$ ($n_0$ being the equilibrium density) and time by the inverse electron plasma frequency $\omega_{p , e}^{-1}=\lambda_{D , e}/v_{th , e}$; consequently, $E$ is scaled by $\sqrt{4\pi n_0 m_e v_{th , e}^2}$ and $f$ by $n_0 / v_{th , e}$.

Eqs.~(\ref{adim1})-(\ref{adim2}), as described before, neglect the motion of protons. However, there is evidence of complex structures in the protons phase space, like in Ref.~\onlinecite{ions}. Nevertheless, the fixed ions assumption is adopted in many recent works, even about 1D-1V simulations \cite{BGK}, in order to explore high-frequency electron dynamics.

Collisions are modeled through the Dougherty operator \cite{Dough1,Dough2}, which, in scaled units, reads:
\begin{equation}\label{dougherty}
\frac{\partial f}{\partial t}\bigr\rvert_{coll}=\nu(n_e,T_e)\frac{\partial}{\partial v}\left[T_e(f)\frac{\partial f}{\partial v}+(v-U_e(f))f\right]
\end{equation}
where $n_e=\int f dv$ is the electron density, $U_e=\int v f dv / n_e$ is the electron bulk speed and $T_e=\int (v-U_e)^2 f dv/n_e$ is the electron temperature; the collision frequency has the form $\nu(n_e,T_e)=\nu_0 n_e / T_e^{3/2}$, where $\nu_0=-g \ln g / \left( 24 \pi \right)$, $g$ being the plasma parameter. This collisional operator conserves mass, energy and momentum and has a generic Maxwellian as the unique equilibrium solution. Details on the numerical approximation of the Dougherty operator can be found in Ref.~\onlinecite{EulColl}.

Note that the Dougherty operator is energy conserving only in the continuous limit on an infinite domain. The discrete form on the truncated velocity domain is conservative only with special treatment \cite{Hakim}. However, we have analyzed further simulations with different values of $L_v$ and we have concluded that the chosen intervals $[-L_v,L_v]$ for runs shown in this work are such that total energy is conserved with relatively small error. For instance, the simulation of the collisional, linear regime with $L_v=6$ is such that the total energy variation is less than $10^{-5} \%$.

The two-dimensional phase space domain $[0,L_x] \times [-L_v,L_v]$ is discretized homogeneously through $N_x \times N_v$ grid points. Periodic boundary conditions are implemented in physical space both for $f$ and $E$, while in velocity space $f$ is assumed to be null for $|v|>L_v$. Eqs.~(\ref{adim1})-(\ref{adim2}) are solved through the method described in Ref.~\onlinecite{EulColl}, based on a third-order finite volume scheme for the advection equations in physical and velocity space \cite{VanLeer,Vl1,Vl2,Vl3,Vl4}, separately. The time evolution of $f$ in phase space is computed by using the original splitting scheme \cite{ChengKnorr} (see also Refs.~\onlinecite{split1,split2}) for time integration in the collisionless case, which is generalized as in Ref.~\onlinecite{VLnh} in the presence of collisions. A standard FFT routine is employed to integrate the Poisson equation for $E$. The system evolution is followed in each numerical run up to a time $T_{max}$ and the time step $\Delta t$ is chosen to be small enough such that Courant-Friedrichs-Levy condition for numerical stability is satisfied \cite{CFL}.

\subsection{Initial conditions}
We performed seven numerical runs with different initial conditions and input parameters, as summarized in Table \ref{tabella}. In this table, we report the box sizes $L_x$ and $L_v$, the number of mesh points $N_x$ and $N_v$, the amplitude of the initial perturbation $A$, and the plasma parameter $g$. In addition, the last column of the Table summarizes the regime of the numerical experiment. 
\begin{table*}
\caption{\label{tabella}Parameters of the simulations. In the $x$ space the simulation box spans from $0$ to $L_x$, comprising $N_x$ mesh points, while in the $v$ space it extends from $-L_v$ to $L_v$, comprising $N_v$ mesh points. $T_{max}$ is the simulation time, $A$ represents the amplitude of the initial perturbation, while $g$ denotes the plasma parameter.}
\begin{ruledtabular}
\begin{tabular}{ccccccccc}
 Run & $L_x$ & $N_x$ & $L_v$ & $N_v$ & $T_{max}$ & $A$ & $g$ & Regime \\
 \hline
 I & $6 \pi$ & 512 & $2 \pi$ & 1921 & 480  & $10^{-4}$ & 0 & Linear damping \\
 II & $6 \pi$ & 512 & $2 \pi$ & 1921 & 480  & $10^{-4}$ & $1.5 \times 10^{-3}$ & Linear damping\\
 III & 18 & 512 & 6 & 12001 & 800  & 0.1 & 0 & Nonlinear trapping \\
 IV & 18 & 512 & 6 & 4001 & 800  & 0.1 & $10^{-3}$ & Nonlinear trapping\\
 V & 20 & 512 & 6 & 4001 & 800 & $10^{-3}$ & $5 \times 10^{-4}$ & Bump-on-tail instability \\
 VI & 64 & 512 & 8 & 6001 & 300 & $10^{-4}$ & $1.5 \times 10^{-3}$ & Two-stream instability\\
 VII & 90 & 1024 & 6 & 6001 & 300 & 0.4 & $10^{-3}$ & Langmuir turbulence\\
\end{tabular}
\end{ruledtabular}
\end{table*}

In Runs I-IV and VII, we assume $f_{eq} = f_{MB}$, while in Run V the equilibrium distribution has a central main beam and two secondary beams and it can be expressed in dimensionless units as:
\begin{align} 
f_{eq} (v) = &\frac{1}{\sqrt{2 \pi}} \frac{a_{1}}{v_{th \; 1}} e^{-\frac{v^2}{2 v_{th \; 1}^2}} +\frac{1}{\sqrt{2 \pi}}\frac{a_{2}}{v_{th \; 2}}\nonumber \\
& \times \left[ e^{-\frac{\left( v - v_{0} \right)^2}{2 v_{th \; 2}^2}} +  e^{-\frac{\left( v + v_{0} \right)^2}{2 v_{th \; 2}^2}} \right], \label{bot}
\end{align}
with $a_1+2a_2 = 1$ ($v_0$, $v_{th \; 1}$ and $v_{th \; 2}$ are scaled by $v_{th , e}$). Finally, in Run VI the distribution function at equilibrium is:
\begin{equation} \label{tb}
f_{eq} (v) = \frac{1}{\sqrt{2 \pi}} \left[ \frac{a_{1}}{v_{th \; 1}} e^{-\frac{\left( v - v_{1} \right)^2}{2 v_{th \; 1}^2}} + \frac{a_{2}}{v_{th \; 2}} e^{-\frac{\left( v - v_{2} \right)^2}{2 v_{th \; 2}^2}} \right] ,
\end{equation}
with $a_1+a_2 = 1$.

Three different forms of the initial perturbation $\delta f$ have been considered:
\begin{subnumcases}{\delta f (x,v,0) = }
A \cos \left( k_0 x \right) f_{eq} (v) , \label{df1}
\\
\frac{A}{3} \sum_{r=0}^{2} \cos \left( 2^r k_0 x \right) f_{eq} (v) , \label{df2}
\\
A k_0 \sum_{r=1}^{10} r \cos \left( r k_0 x + \phi_r \right) f_{eq} (v) , \label{df3}
\end{subnumcases}
where $\phi_r$'s are random phases. Perturbations in Eq.~\eqref{df1} have been used for Runs I-V, while those in Eqs.~\eqref{df2} and \eqref{df3} have been used for Runs VI and VII, respectively.

\subsection{The Fourier-Hermite analysis} \label{FH_analysis}
Once one has discretized phase space and time, the dimensionless distribution function $f$ can be analyzed through the Fourier-Hermite decomposition. 

Hermite polynomials form a well-known class of functions which have many relevant applications in mathematics and physics. Given an integer $m \geq 0$, the $m$-th Hermite polynomial is $H_m(v)=(-1)^m e^{v^2} (d^m / dv^m) e^{-v^2}$. The Hermite class $\lbrace H_m \rbrace$ is defined in such a way to form an orthogonal basis in $\mathbb{R}$ with weight $e^{-v^2}$. Starting from this property, one can build the $m$-th Gauss-Hermite function
\begin{equation}
\psi_m(v)=\frac{H_m (v)}{\sqrt{2^mm!\sqrt{\pi}}}e^{-\frac{v^2}{2}},
\label{eq:hmm}
\end{equation}
and can get an orthonormal basis where $\int_{-\infty}^{\infty} \psi_m(v) \psi_{m'}(v) dv = \delta_{m \; m'}$. In this way, one can decompose the distribution function in the velocity space as $f(x,v,t) = \sum_{m=0}^{\infty} f_m (x,t) \psi_m (v)$ and get the Hermite coefficients
\begin{equation}
f_m (x,t) = \int_{-\infty}^{+\infty} f(x, v, t) \psi_m(v) dv.
\label{eq:proj}
\end{equation}
In dimensionless units $f_{MB} = \exp (-v^2 / 2) / \sqrt{2 \pi}$, so the Hermite coefficient $f_0$, as it can be evinced from Eq.~(\ref{eq:hmm}), contains the Maxwellian part of $f$. On the other hand, high-order coefficients represent deviations from $f_{MB}$. If $m \gg 1$ it is possible to show that $\psi_m (v) \sim C_m \cos \left( \sqrt{2 m} v - m \pi / 2 \right)$ (see Appendix \ref{AppA}), so fluctuations with period $\delta v$ are associated with Hermite modes such that $m \approx 2 \pi^2 / \delta v^2$.

The integral in Eq.~(\ref{eq:proj}) has been obtained by means of an adequate quadrature. In order to avoid spurious aliasing and convergence problems, the Gauss-Hermite quadrature has been used \cite{Gauss} (see also Refs.~\onlinecite{MMS, Pezziquad}). In particular, Eq.~\eqref{eq:proj} is equivalent to
\begin{equation}
f_m (x,t) = \int_{-\infty}^{+\infty} g_m(x,v,t) W(v) dv
\end{equation}
where $g_m(v) = f (v) H_m (v) e^{v^2 / 2} / \sqrt{2^mm!\sqrt{\pi}}$ and $W(v)=e^{-v^2}$. Starting from the discretized values of the distribution function $f(x_j, v_l, t_p)$, the Gauss-Hermite quadrature permits to write
\begin{equation}
f_m (x_j,t_p) \approx \sum_{q=1}^{M} w_q g_m(x_j, v_q, t_p),
\end{equation}
where $v_q$'s are the roots of the Hermite polynomial $H_M(v)$ (large values of $M$ increase approximation accuracy) and $w_q$'s are weights evaluated as in Ref.~\onlinecite{quad}:
\begin{equation}
w_q=\frac{2^{M-1}M!\sqrt{\pi}}{M^2H^2_{M-1}(v_q)} .
\end{equation}
In our analysis, Hermite coefficients are given as:
\begin{equation}
f_m (x_j,t_p) = \sum_{v_q \in [-L_v,L_v]} w_q g_m (x_j,v_q,t_p) ,
\label{eq:appr}
\end{equation}
with $0 \leq m < M$. As the function $g_m$ is defined at $\lbrace v_l \rbrace$ points, a standard bilinear interpolation has been used to estimate it on $\lbrace v_q \rbrace$ grid. We set $M = 800$ for each simulation, this meaning that the not-equally spaced grid for the Gauss-Hermite quadrature procedure is composed of the roots of the 800-th Hermite polynomial, in the interval $[-L_v,L_v]$.

Finally, we analyze the spatial perturbations of the distribution function by performing a classical, optimized fast Fourier transform (FFT) algorithm of the coefficients of Eq.~(\ref{eq:appr}):
\begin{equation}
f_{n \; m} (t_p) = \sum_{j=0}^{N_x - 1} f_m (x_j,t_p) e^{-i k_n x_j} ,
\end{equation}
where $n$ is the index referred to the $n$-th Fourier mode with $k_n = n k_0 = 2 \pi n / L_x$.

\subsection{Free energy conservation law} \label{en_cons}
The collisional VP system conserves total energy, as discussed in detail in Appendix \ref{AppB}. However, we expect, as time passes, that the free energy associated with the system decreases because of the collisional term, with an increase of entropy and thermal energy. This conversion rate can be deduced by the following conservation law, written in dimensionless units:
\begin{align}
&\frac{d}{dt} \left[ \int \frac{E^2}{2} dx + \int f \ln \left( \frac{f}{f_{MB}} \right) dxdv \right] \nonumber \\
& \qquad = \int \ln \left( \frac{f}{f_{MB}} \right) \frac{\partial f}{\partial t} \bigr\rvert_{coll} dxdv . \label{ECL}
\end{align}
The left-hand side of Eq.~\eqref{ECL} is the time derivative of the free energy of the system, understood as the effective energy that it can use to do work. In fact, the first integral in square brackets is the electric energy $\mathcal{E}$, while the second one is the thermodynamic free energy of particles $\mathcal{F}$: this can be deduced by noticing that
\begin{align}
\mathcal{F} &= \int \frac{(v-U_e)^2}{2} f dxdv + \int \frac{U_e^2}{2} f dxdv + \int f \ln f dxdv \nonumber \\
&= U + K - S ,
\end{align}
where $U$ is the internal energy, $K$ is the bulk kinetic energy and $S$ is the entropy of the plasma. The right-hand side, which for the sake of simplicity will be indicated as $\mathdutchcal{c}$, expresses the role of $\partial f / \partial t \rvert_{coll}$ in the thermal dissipation. For further details, see Appendix \ref{AppB}. Observing the conservation law, we expect that in the collisionless case there is a free energy exchange between the electric field and equilibrium perturbations. On the other hand, when collisions are turned on, both tend to be damped until $f = f_{MB}$, when the right-hand side of Eq.~\eqref{ECL} vanishes and the free energy reaches its minimum, i.e. it becomes null.

This conservation law, which can be seen as a special case of the law driven in Ref.~\onlinecite{Cassak}, has an approximate form reported in Refs.~\onlinecite{MixAdv, Budget}. It can be obtained from Eq.~\eqref{ECL} through the Taylor expansion of $\mathcal{F}$ and $\mathdutchcal{c}$, calculated by setting $f = f_{MB} + \Delta f$ (note: $\Delta f = \delta f$ if and only if $f_{eq} = f_{MB}$). This gives $\mathcal{F} = \sum_{\beta = 1}^{\infty} \mathcal{F}_\beta$ and $\mathdutchcal{c} = \sum_{\beta = 1}^{\infty} \mathdutchcal{c}_\beta$, where:
\usetagform{nowidth}
\begin{subequations}
\begin{gather}
\mathcal{F}_\beta(t) = \frac{(-1)^{\beta+1}}{\beta(\beta+1)} \int \frac{\Delta f^{\beta+1}(x,v,t)}{f_{MB}^\beta(v)} dxdv ,
\\[5pt] 
\begin{align}
\mathdutchcal{c}_\beta(t) =& \frac{(-1)^{\beta+1}}{\beta} \int \left( \frac{\Delta f(x,v,t)}{f_{MB}(v)} \right)^\beta \nonumber 
\\
& \times
\frac{\partial f}{\partial t} \bigr\rvert_{coll} (x,v,t) dxdv . 
\end{align}
\end{gather}
\end{subequations}
\usetagform{default}
The approximation of Refs.~\onlinecite{MixAdv, Budget} for the 1D-1V VP system is given by the truncation of the series at $\beta = 1$.  

In each numerical run, we implement the integrated form of the conservation law. Its exact version is 
\begin{equation} 
\label{budget_int}
\mathcal{E} (t) + \mathcal{F} (t) - \mathcal{C} (t) = C_0 ,
\end{equation}
where $\mathcal{C} (t) = \int_{0}^{t} \mathdutchcal{c} (t') dt'$ and $C_0 = \mathcal{E} (0) + \mathcal{F} (0)$ is imposed by initial conditions. Instead, by writing $\mathcal{C} (t) = \sum_{\beta = 1}^{\infty} \mathcal{C}_{\beta} (t) = \sum_{\beta = 1}^{\infty} \int_{0}^{t} \mathdutchcal{c}_{\beta} (t') dt'$, the approximated version appears as
\begin{equation} 
\label{budget_int_appr}
\mathcal{E} (t) + \mathcal{F}_1 (t) - \mathcal{C}_1 (t) \approx \mathcal{E} (0) + \mathcal{F}_1 (0) .
\end{equation}
More specifically, we compute the terms of the left-hand side of Eqs.~(\ref{budget_int})-(\ref{budget_int_appr}) with a sampling of $10 \Delta t$. In this way, we can analyze the time evolution of the exact terms, in order to get insights into the energy exchanges mentioned above and point out the role of collisions in the simulated regimes. At the same time, we can compare $\mathcal{F}$ with $\mathcal{F}_1$ and $\mathcal{C}$ with $\mathcal{C}_1$: this permits us to observe the behavior of the approximation of Refs.~\onlinecite{MixAdv, Budget} when it is applied to systems which are very far from the thermodynamic equilibrium, and quantify the relevance of high-order terms of $\mathcal{F}$ and $\mathcal{C}$ as time passes. 

\begin{figure*}
   \begin{center}
        \includegraphics[scale=0.96]{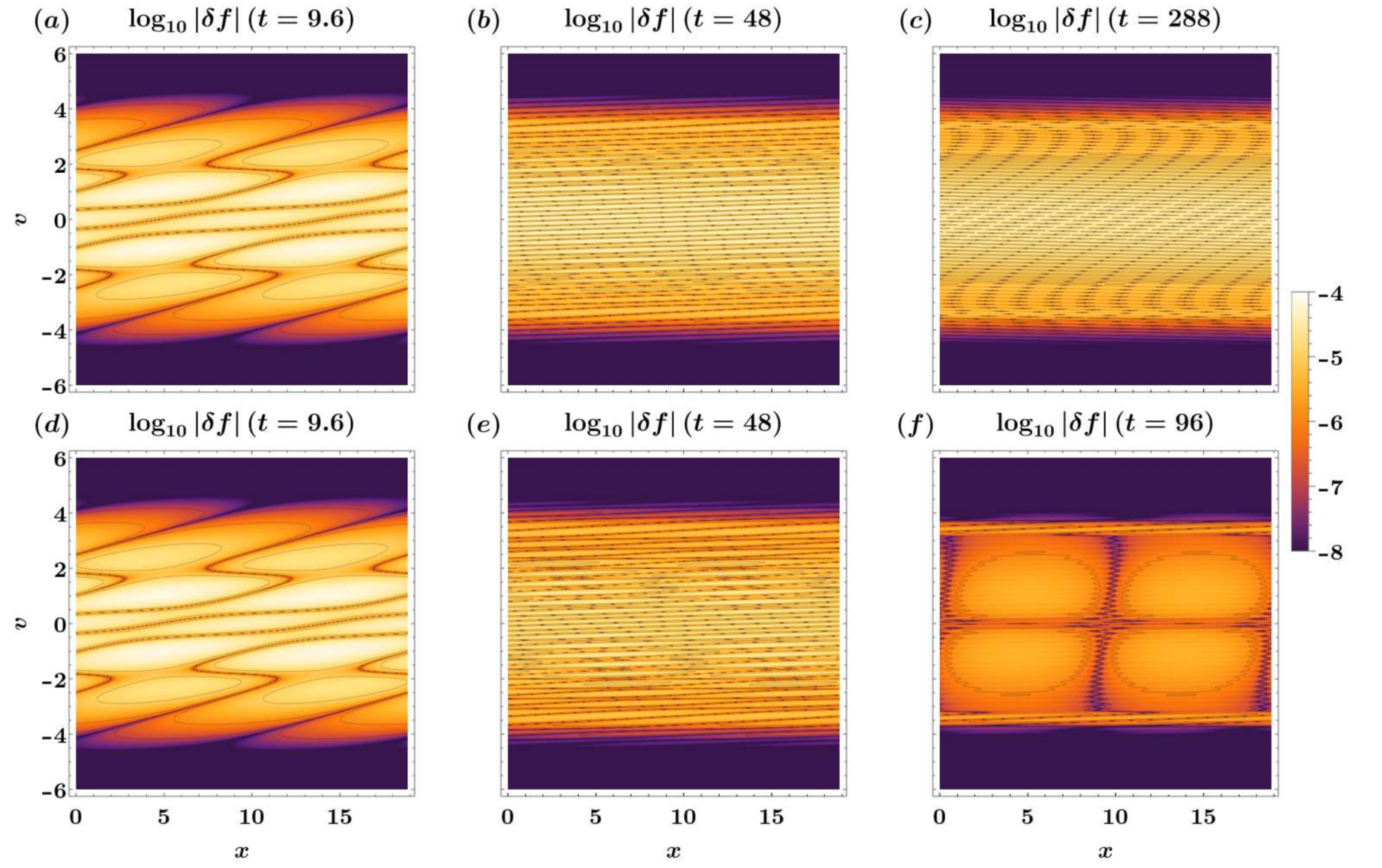}
        \caption{Phase space contours of the perturbed distribution function $\delta f$ at different times for Run I (collisionless linear damping) [panels (a)-(c)] and Run II (collisional linear damping) [panels (d)-(f)].}
        \label{fig2}
    \end{center}
\end{figure*}

\section{Numerical results} \label{secIII}
We now discuss the direct numerical simulations summarized in Table \ref{tabella}. In particular, we begin with an overview of each run, concentrating on the time evolution of the distribution function in phase space. The analysis is then discussed in terms of the distribution function's Hermite and Fourier-Hermite spectra, their relationship with the electric field spectra, and free energy transfer across various spatial and velocity scales. Finally, we examine the free energy budget, as described in Sec.~\ref{en_cons}.

\begin{figure*}
   \begin{center}
        \includegraphics[scale=0.96]{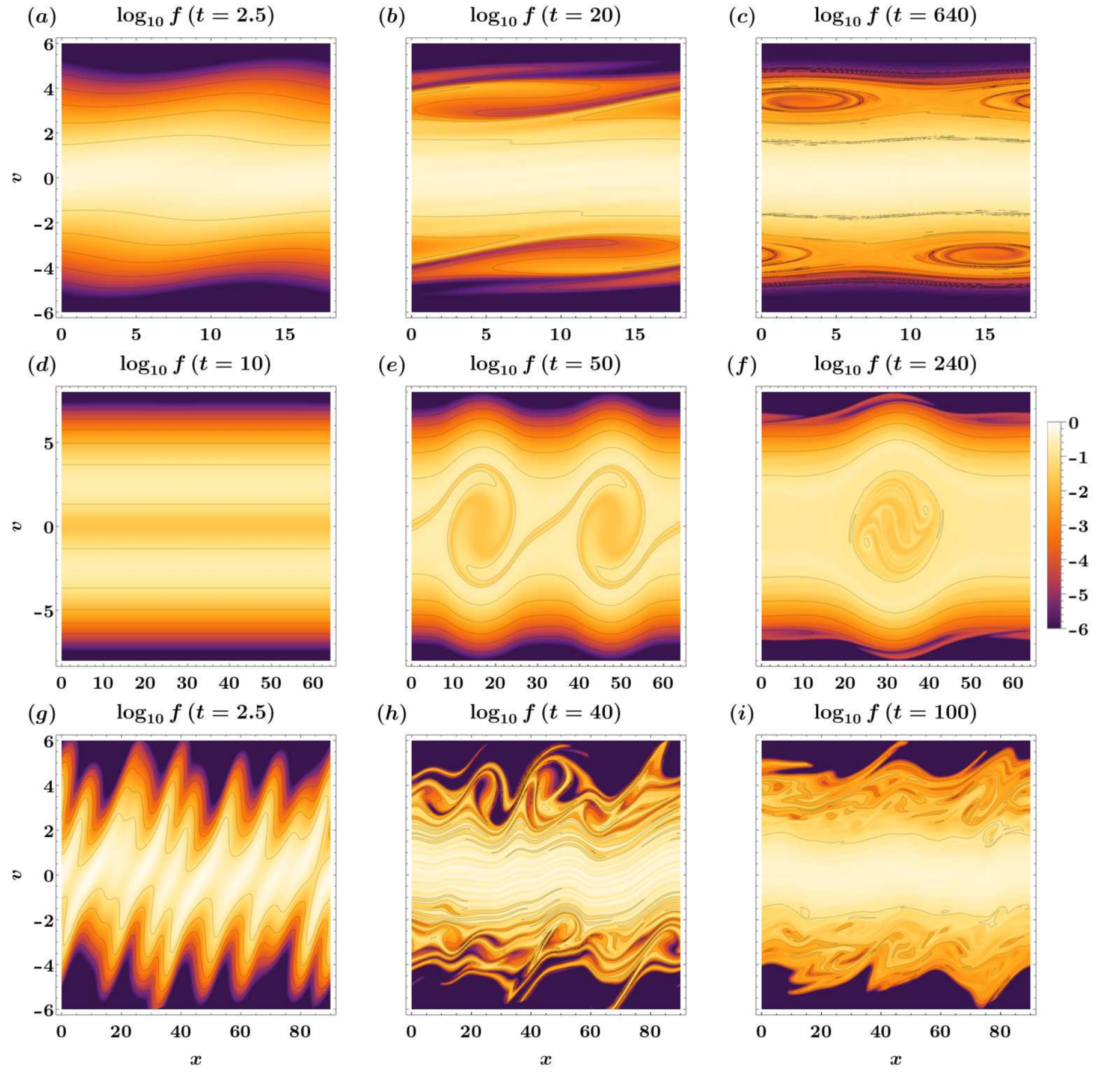}
        \caption{Phase space contours of the distribution function $f$ at different times for Run III (collisionless nonlinear trapping) [panels (a)-(c)], Run VI (two-stream instability) [panels (d)-(f)], and Run VII (Langmuir turbulence) [panels (g)-(i)].}
        \label{fig3}
    \end{center}
\end{figure*}

In Runs I and II, we reproduce numerically the response of the system to a small amplitude monochromatic perturbation imposed on a Maxwellian equilibrium, without and with collisions, respectively. For times shorter than the trapping time \cite{ONeil}, that is $\tau = 2 \pi / \sqrt{A} \simeq 630$, the electric perturbation is Landau-damped in time \cite{Landau}. In fact, the electric energy clearly follows the trend $e^{2 \gamma t}$, where $\gamma \simeq -2.57 \times 10^{-2}$ denotes the imaginary part of the dispersion relation of Langmuir waves: $\gamma$ is estimated by calculating numerically the Landau integral in the complex plane. At the same time, phase space filamentation occurs in the perturbed distribution function \cite{Langmuir}. This is visible in Fig.~\ref{fig2}, where the phase space portrait of $\delta f$ is shown at different times, for Run I [panels (a)-(c)] and Run II [panels (d)-(f)]. As expected from the linear theory, in the collisionless run, small velocity scales are generated as the system evolves, while in Run II the effect of weak collisions appears evident at times $t > 50$, damping out small velocity scales [panel (f)] and working to restore local thermodynamic equilibrium.

Since in Runs I and II the initial perturbation is very small and no instability is triggered, $E \approx 0$ throughout the entire simulations. This means that $f$ is near the solution of the Vlasov equation for truly free streaming particles, i.e. in Run I (and in Run II too, before the dominance of collisions) $f \approx A \exp [i k_0 (x - v t)] f_{eq}$. If the mesh spacing in velocity space is $\Delta v$, there is a numerical recurrence occurring at $T_R = 2\pi/(k\Delta v)$, which can be prevented by a sufficiently high value of $g$ (see Ref.~\onlinecite{NumRec}). In any case, in our linear simulations the recurrence is avoided because for both Runs I and II $T_R \simeq 2900$ is much larger than $T_{max}$.

When $T_{max}$ is of the order (or larger) than $\tau$, nonlinear effects come into play in the process of wave-particle resonant interaction and particles are trapped in the wave potential well. As a consequence, the distribution function appears significantly distorted in the resonant region (in the range of velocities close to the wave phase speed), with the generation of phase space vortices -- the smoking gun of trapping processes. This is visible in Fig.~\ref{fig3} (a)-(c), where the contours of the distribution function in phase space are reported at three different times for the collisionless Run III ($\tau \simeq 20$). Two counter-propagating trapping structures are clearly visible at late times [panel (c)], and when weak collisions are present (Run IV, not shown here), phase space vortices are dissipated in the long time limit.

In Run V, the initial equilibrium in Eq.~(\ref{bot}) is perturbed in such a manner that the bump-on-tail instability is triggered. We set $a_1 = 0.98$, $a_2 = 0.01$, $v_{th \; 1} = 1$, $v_{th \; 2} = \sqrt{0.2}$ and $v_0 = v_{0}^{max} \simeq 4.133$, where $v_{0}^{max}$ has been chosen to maximize the linear growth rate of the instability \cite{bump}. More specifically, the Landau integral for this simulation gives $\gamma \simeq 5.79 \times 10^{-2}$.

The two-stream instability has been simulated in Run VI, in which the initial equilibrium consisting of two counter-propagating electron populations with the same density has been perturbed through the initial disturbance in Eq.~(\ref{df2}). In particular, at $t = 0$, we set $a_1 = a_2 = 1 / 2$, $v_{th \; 1} = v_{th \; 2} = 1$ and $v_1 = -v_2 > 0$. Under these conditions, according to Refs.~\onlinecite{beam,tutorial,TwoBeam}, a small perturbation in $f$ with wave number $k$ is unstable when $k < 1 / v_1$. In our case $v_1 = 2.5$ and the Fourier modes excited at $t = 0$ are $k_0$, $2k_0$ and $4k_0$ (with $k_0 = \pi / 32$). All of these modes are evidently unstable, with $k=2k_0$ which appears to be the dominant one.

The phase space evolution of the electron distribution function is shown in Fig.~\ref{fig3} (d)-(f), at different times. Two vortical structures [panel (d)] are clearly generated in phase space (corresponding to the most unstable Fourier mode), which merge at later times [panel (f)], resulting in a stationary configuration with a single persistent vortex. This is a clear indication of a process of decay to a longer wavelength mode, triggered by the general tendency of trapped particle vortices to coalesce \cite{vortex1,vortex2,vortex3,vortex4,vortex5,vortex6}. We point out that at $t=240$ [panel (f)] weak collisionality has not significantly affected the phase space distribution yet.

Finally, in Run VII we focus on the simulation of Langmuir turbulence in the presence of collisions; the initial Maxwellian equilibrium ($f_{eq} = f_{MB}$) is perturbed through a superposition of large amplitude Fourier modes (the first 10 Fourier modes) with random phases, set in such a way that the corresponding Fourier modes all have the same energy. The amplitude $A=0.4$ of the excited Fourier harmonics has been chosen in such a way that the trapping time associated with each mode is much smaller than $T_{max}$. This ensures that all excited modes quickly undergo from linear to nonlinear regimes, triggering an efficient energy transfer towards small spatial scales. The excitation of short wavelength modes through this energy cascade is associated with the generation of fine velocity structures in the particle distribution function, which are eventually smoothed out by collisional effects. In Fig.~\ref{fig3} (g)-(i), we report the contour plot of the distribution function in phase space at three different times, where deformations and distortions due to nonlinear effects are clearly visible at $t=40$ [panel (h)] and the effect of dissipation due to collisions occurs at larger times [panel (i)]. This case manifests a very good example of fully developed, homogeneous phase space turbulence.

\subsection{Hermite spectra} \label{Her}
\begin{figure*}
   \begin{center}
        \includegraphics[scale=0.96]{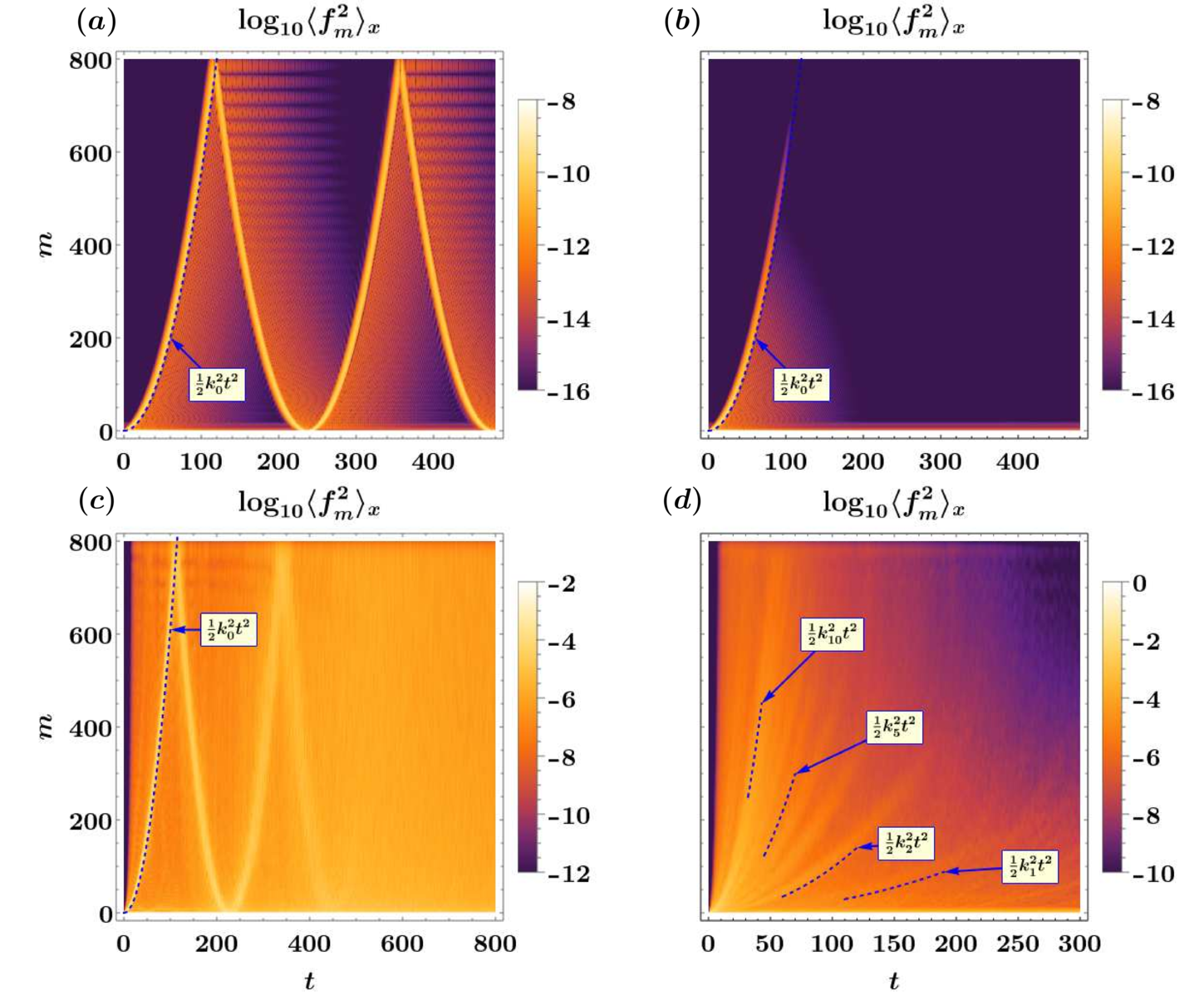} 
        \caption{Spatially averaged Hermite spectrum $\langle f_{m}^2 \rangle_x$ as a function of $m$ and $t$ for Run I (collisionless linear damping) [panel (a)], Run II (collisional linear damping) [panel (b)], Run III (collisionless nonlinear trapping) [panel (c)], and Run VII (Langmuir turbulence) [panel (d)]. The dashed blue lines in each plot represent the curves $m^\star (t) = k_0^2 t^2 / 2$ in Runs I-III and $m^\star_{n} (t) = k_n^2 t^2 / 2$ for some $n$ in Run VII.}
        \label{fig4}
    \end{center}
\end{figure*}

\begin{figure*}
   \begin{center}
        \includegraphics[scale=0.96]{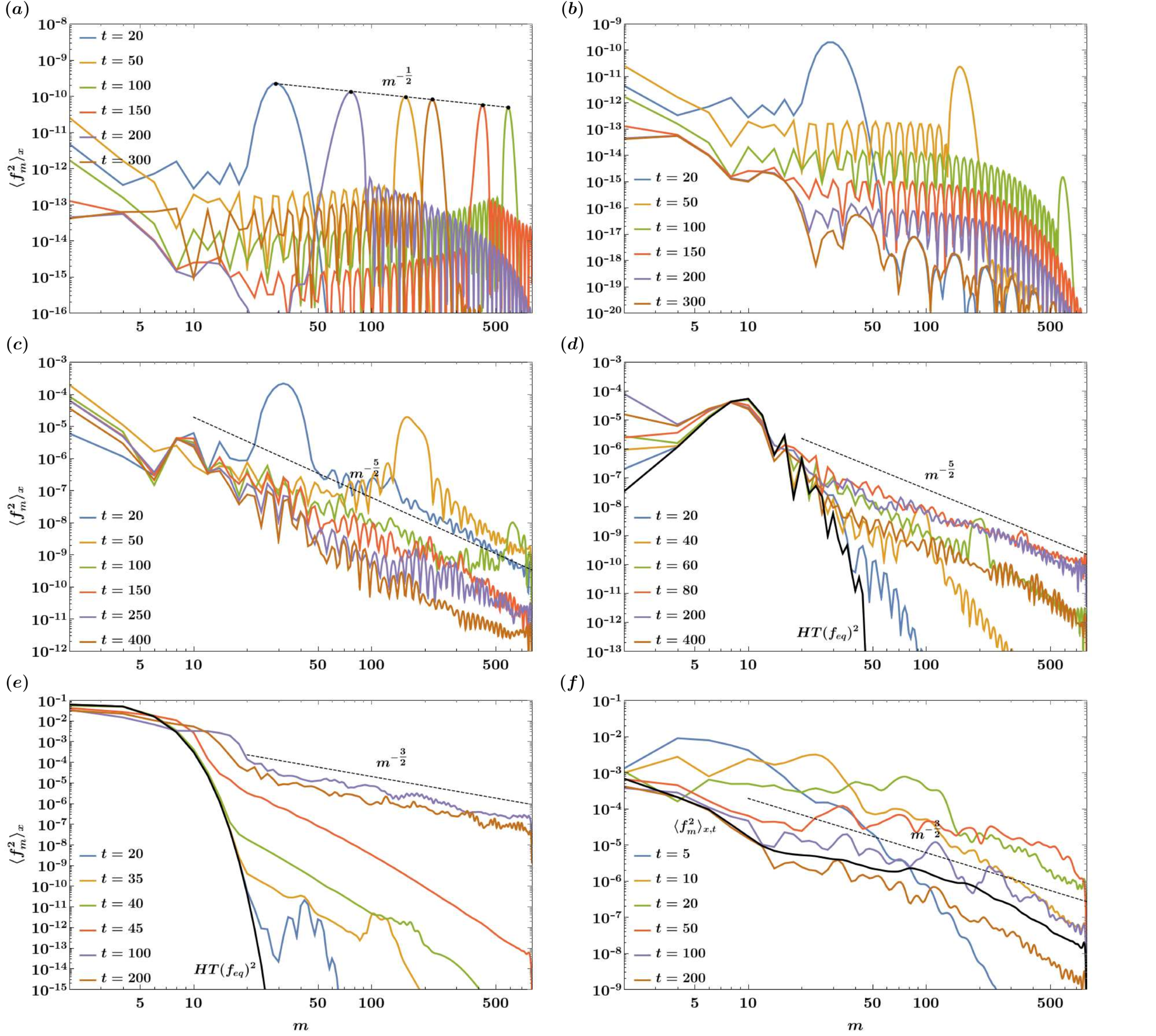} 
        \caption{Cuts of spectra $\langle f_{m}^2 \rangle_x$ for fixed values of $t$ in Run I (collisionless linear damping) [panel (a)], Run II (collisional linear damping) [panel (b)], Run IV (collisional nonlinear trapping) [panel (c)], Run V (bump-on-tail instability) [panel (d)], Run VI (two-stream instability) [panel (e)], and Run VII (Langmuir turbulence) [panel (f)]. Dashed black lines: comparisons of power laws $m^{-1/2}$ [panel (a)], $m^{-5/2}$ [panels (c)-(d)], and $m^{-3/2}$ [panels (e)-(f)]. Solid black lines: Hermite transform of the equilibrium distribution [panels (d)-(e)] and time average $\langle f_{m}^2 \rangle_{x,t}$ for $t \in [80,300]$ [panel (f)].}
        \label{fig5}
    \end{center}
\end{figure*}

\begin{figure*}
   \begin{center}
        \includegraphics[scale=0.96]{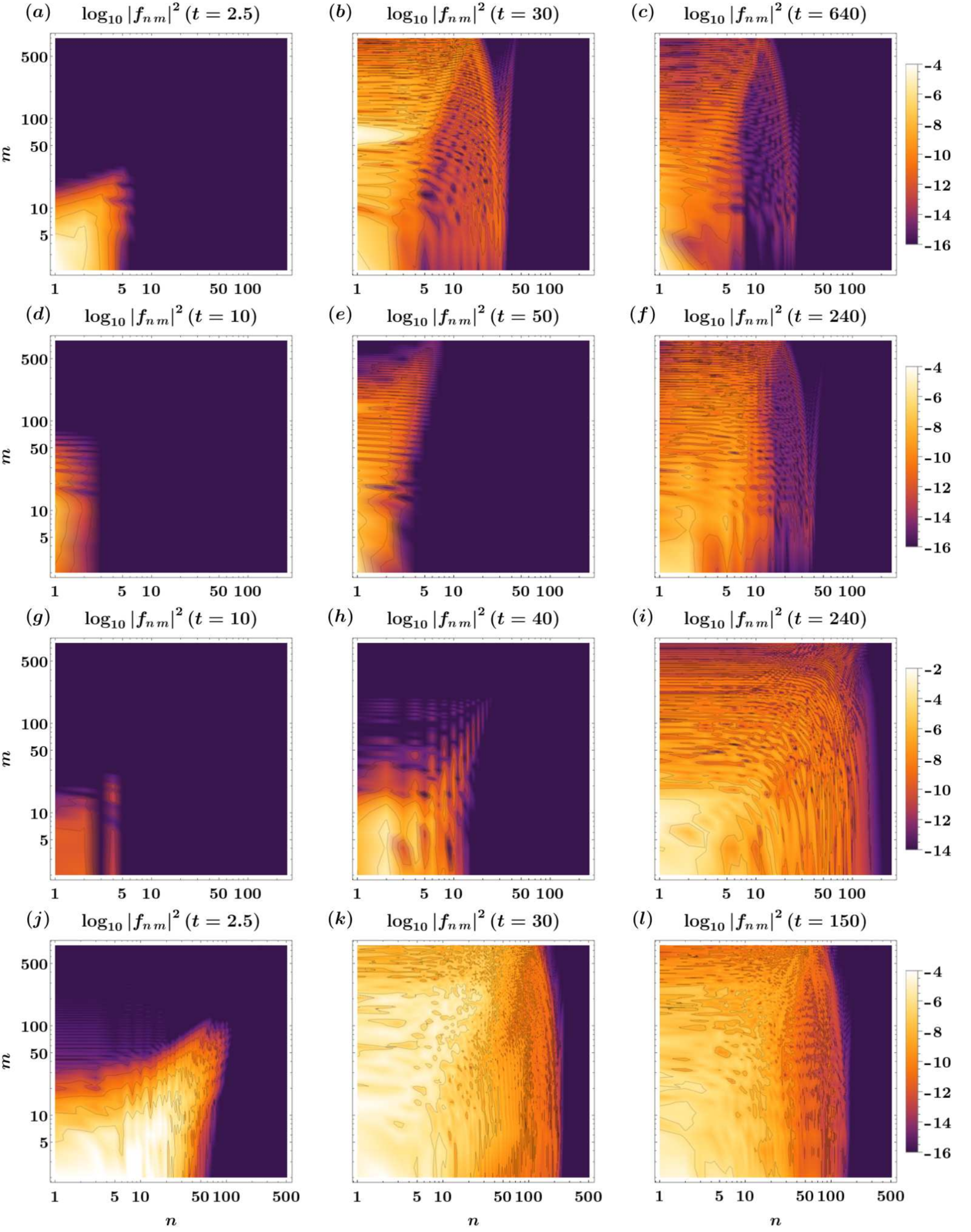}
        \caption{Contour plots of Fourier-Hermite coefficients $f_{n \; m}$ at different times for Run IV (collisional nonlinear trapping) [panels (a)-(c)], Run V (bump-on-tail instability) [panels (d)-(f)], Run VI (two-stream instability) [panels (g)-(i)], and Run VII (Langmuir turbulence) [panels (j)-(l)].}
        \label{fig6}
    \end{center}
\end{figure*}

The analysis based on the Hermite transform of the distribution function, as discussed in Sec.~\ref{FH_analysis}, allows us to emphasize the mechanism of generation of small velocity scales due to kinetic effects and its interplay and competition with collisions. In fact, as kinetic processes tend to generate fine velocity scales (high $m$ Hermite harmonics), driving locally the distribution function away from thermodynamic equilibrium, the Dougherty operator acts as a diffusion operator, damping mainly the high $m$ Hermite modes, with the asymptotic tendency to thermal equilibrium, where only the Hermite mode with $m = 0$ remains excited \cite{NumRec}. Apparently, in our simulations both an inverse and a direct cascade might take place. For instance, this could happen in Run VI, during the merging of the vortices of Fig.~\ref{fig3} (e). We stress that we are not quantitatively measuring the flux direction here. However, the concept of inverse cascade is very likely to occur (see also Refs.~\onlinecite{casc1, casc2}).

In Fig.~\ref{fig4}, the contours of the spatially averaged Hermite spectrum $\langle f_{m}^2 \rangle_x$ of the particle distribution are shown in the $(t,m)$ plane, for Runs I-III [panels (a)-(c)] and Run VII [panel (d)]. Panels (a) and (b) refer to simulations of linear damping in the absence and presence of collisions, respectively. We observe that, in both cases, at a given time enstrophy is highly concentrated in a few Hermite modes and flows in time towards higher $m$'s, along a parabolic path. In fact, until the approximation $f \approx A \exp [i k_0 (x - v t)] f_{eq}$ is invalidated by collisional effects, it is trivial to deduce that the typical velocity scale of $\delta f$ the perturbation in the particle distribution depends on time as $\delta v (t) = 2 \pi / (k_0 t)$. This corresponds, as explained in Sec.~\ref{FH_analysis}, to an excitation of the Hermite mode with $m \approx 2 \pi^2 / \delta v^2$ regardless of the expression of $C_m$; then, one gets the time evolution of the Hermite spectrum as $m^\star (t) \approx k_0^2 t^2 / 2$ (the expected parabolic time evolutions of the spectra are reported as dashed blue lines in all panels of Fig.~\ref{fig4}).

As it is evident from panel (a) of Fig.~\ref{fig4}, the enstrophy flows towards higher Hermite modes until it reaches the $800$-th mode (the maximum mode considered in the analysis) at $t \simeq 120$ and then bounces back. Then the flow tends to periodically bounce back and forth. This is of course a numerical effect due to the finite number of Hermite modes considered in our method. On the other hand, when collisions are present [panel (b)], high Hermite modes are rapidly smoothed out and the first artificial bouncing is thus prevented. For Run III [panel (c)], the behavior is similar to that observed in Run I, except that here nonlinear effects tend to dominate the dynamics and the linear filamentation is less visible in the Hermite spectrum, until it is hidden by the nonlinear spread. Finally, in the turbulent Run VII many Fourier modes are present, interacting nonlinearly with each other, so enstrophy towards high $m$'s flows along different parabolic paths, until nonlinearity tends to distribute it to all available modes and collisional effects become noticeable.

In Fig.~\ref{fig5}, we report the dependence on $m$ of the averaged spectrum $\langle f_{m}^2 \rangle_x$, at fixed time instants; going from panel (a) to panel (f), results for Runs I, II, and IV-VII are displayed. In panel (a), we notice that the peaks in the Hermite spectrum move towards high $m$'s in time and the free energy content of these peaks scales as $m^{-1/2}$ (dashed line); this is due to the asymptotic trend $C_m \sim \sqrt[4]{2 / \left( m \pi^2  \right)}$, as thoroughly discussed in Appendix \ref{AppA}. Run II [panel (b)] is the same as Run I, except that now collisional effects inhibit the development of the Hermite cascade and the $m^{-1/2}$ scaling is no longer recovered.

Panels (c) and (d) display the spectra for the weakly collisional simulations of nonlinear trapping (Run IV) and bump-on-tail instability (Run V); for the latter we also report, for comparison, the HT of $f_{eq}$ as a solid black line. It is worth noting that for Run V the instability leads rapidly the system towards a nonlinear regime, dominated by particle trapping, similarly to Run IV. For both runs, the Hermite spectrum fills in the range of high $m$'s as time goes on, up to the time when kinetic and collisional effects compete and balance each other; the spectral free energy displays a scaling close to $m^{-5/2}$ (dashed lines), confirming the nonlinear trend analytically obtained by Schekochihin \textit{et al.} in Ref.~\onlinecite{MixAdv}.

For Run VI [panel (e)] and Run VII [panel (f)], fully developed Hermite spectra are again visible, with characteristic $m^{-3/2}$ scaling, where evidently the system undergoes a fully nonlinear phase space cascade with arguments similar to the fluidlike Kolmogorov cascade \cite{MMS}. For Run VI, as in the previous panel, the black line is the HT of the two-stream equilibrium distribution. Instead, the black line in panel (f) represents the time average of $\langle f_{m}^2 \rangle_x$ in an interval ($t \in [80,300]$) where the enstrophy saturation at high $m$'s due to the limitation $m \leq 800$ is enough suppressed by the collisional term. This average smooths out the oscillations due to the cascade of the initially excited Fourier modes and better emphasizes the observed power law. 

\subsection{Fourier-Hermite spectra} \label{Fou}

More complete information on the nonlinear system dynamics and on the interplay between kinetic effects and particle collisions can be obtained by taking the Fourier-Hermite transform of the particle distribution function to get $f_{n \; m}(t)$, as discussed in detail in Sec.~\ref{FH_analysis}. The contour plots of $f_{n \; m}$ are shown at different times in Fig.~\ref{fig6}.

Run IV (nonlinear trapping) and Run V (bump-on-tail instability) are characterized by different initial equilibrium distributions (Maxwellian in the first case and Maxwellian with small bumps in the tails in the second one) and by about same values of $g$ (collisionality). The Fourier-Hermite spectra look somewhat different at early times [panels (a)-(b) and (d)-(e)], due essentially to the fact that the initial Hermite spectrum is different in the two cases. Moreover, for Run V, FHT emphasizes the gradual transition from the linear to the nonlinear regime: indeed, at short times [$t = 10$, panel (d)], free energy flows towards high $m$ modes but not towards high $n$ modes, a sign that linear filamentation is at work exciting high $m$'s, but only a single Fourier mode $n$ is present. As time goes on, the instability drives the system to the nonlinear regime, characterized by trapping as in Run IV, and, as a consequence, at large times [panels (c) and (f)], spectra look very similar. Here, nonlinear trapping dominates the dynamics in both cases and the effect of collisions is comparable. It is worth pointing out that both in panels (c) and (f) free energy has developed towards both high $m$'s and $n$'s, clear evidence of a cascade in phase space associated with nonlinear wave-particle interaction.

In Run VI (two-stream instability), whose results are reported in panels (g)-(i), the Fourier-Hermite spectrum at different times reveals a quite rich phenomenology, where both high $n$ and $m$ modes are rapidly excited during the evolution of the instability and fill the whole spectrum at late times [$t = 240$, panel (i)]. The turbulent Run VII in panels (j)-(l) shows a similar behavior of the Hermite spectrum, with an impulsive development of the free energy cascade both along $n$ and $m$ modes, already visible at early times [$t=30$, panel (k)]; moreover, by comparing panel (l) with panel (k), one can notice the effect of collisions in damping out high $n$ modes and, at the same time, decreasing the enstrophy content of high $m$ components.

Spectra displayed in Fig.~\ref{fig6} clearly shows how the Fourier-Hermite analysis performed on the simulations discussed above allows to clearly point out the development of a cascade of free energy in phase space. Thanks to the application of FHT, it is possible to effectively observe how the cascade generally spreads more quickly towards the highest $m$'s rather than the highest $n$, with an anisotropic pattern highly dependent on the plasma regime considered. At the same time, FHT can quantify how the free energy flow is mitigated by the diffusive process triggered by particle collisions.
\subsection{Electric energy spectra} \label{elec}
\begin{figure*}
   \begin{center}
        \includegraphics[scale=0.96]{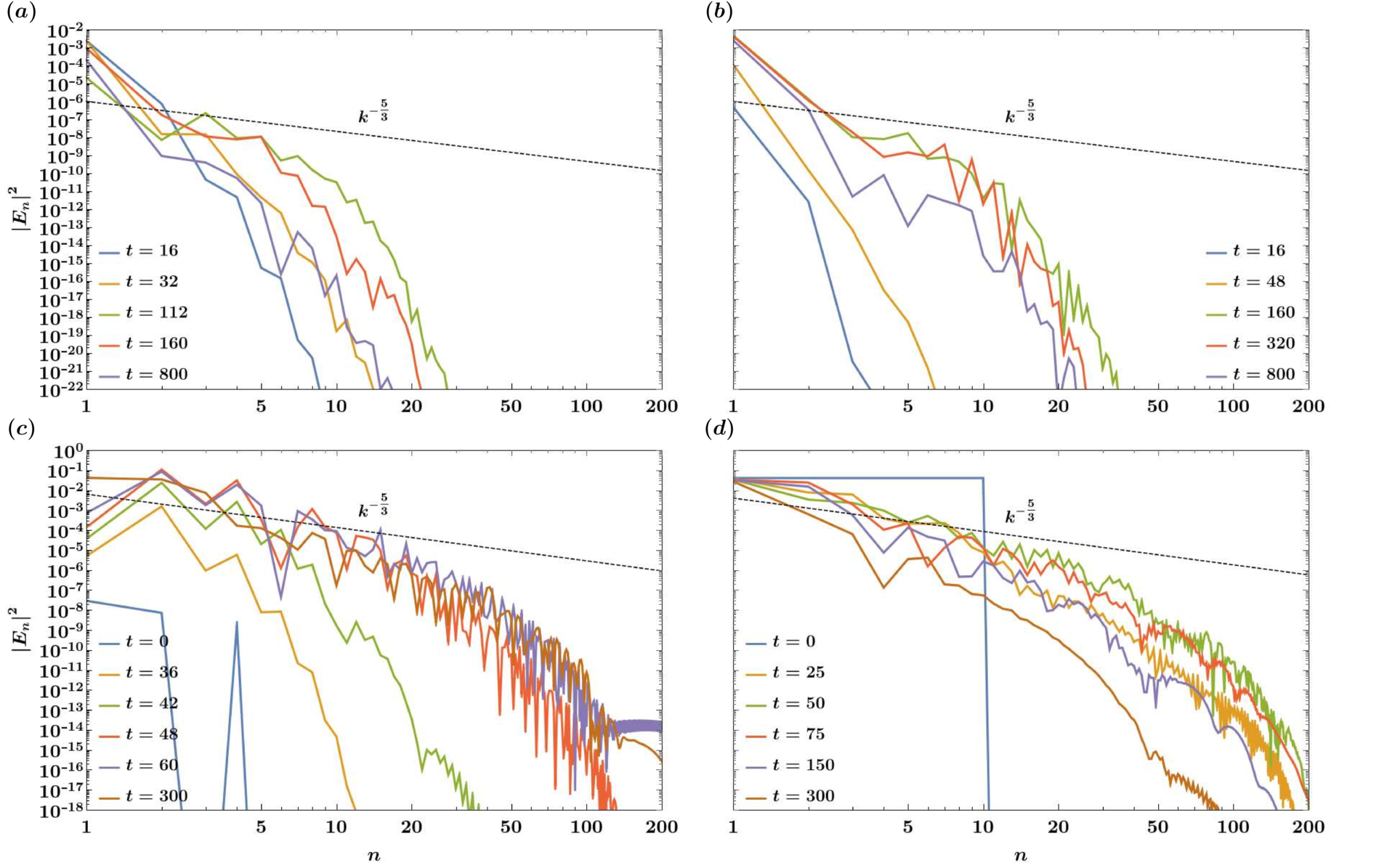} 
        \caption{Smoothed Fourier spectra of the electric field $E_n$ at different times in Run IV (collisional nonlinear trapping) [panel (a)], Run V (bump-on-tail instability) [panel (b)], Run VI (two-stream instability) [panel (c)], and Run VII (Langmuir turbulence) [panel (d)]. Dashed black lines: Kolmogorov scaling $k^{-5/3}$ used as reference.}
        \label{fig7}
    \end{center}
\end{figure*}

Important complementary insights into the nature of wave-particle interaction can be gained by looking at the Fourier spectra of the electric signals from the simulations. In Fig.~\ref{fig7}, we report the spectral electric energy as a function of the Fourier mode number $n$ at different times, for Runs IV-VII. Spectra are displayed after a smoothing procedure based on the moving average technique, in order to improve the clarity of their trend. The Kolmogorov scaling $k^{-5/3}$ expected for stationary, homogeneous, and isotropic turbulence \cite{Kol} is reported as a benchmark in all panels as a dashed black line. From the direct comparison, one can realize that in the two-stream instability [panel (c)] and Langmuir turbulence simulations [panel (d)] the energy transfer towards the tail of the spectrum is much more efficient than in the case of nonlinear trapping and bump-on-tail instability. This is mainly because both in Runs IV and V a single Fourier mode is perturbed at $t = 0$ and remains dominant at all times; here, nonlinear interactions and couplings between Fourier modes are somewhat inhibited. In Runs VI and VII, instead, the cascade is favored by the fact that several Fourier modes are excited in the initial perturbation and can interact nonlinearly in a very efficient way.

Another important aspect one can notice in all panels of Fig.~\ref{fig7} is that collisions come into play at late times (when many Fourier harmonics are excited) because higher modes tend to be suppressed quicker than the lower ones. This effect can be understood by taking into account that the Dougherty operator involves velocity gradients of the particle distribution function and does not play any role when the plasma is at thermal equilibrium. This means that collisional processes become efficient only once kinetic effects have produced distortions (sharp velocity gradients) and local departures from Maxwellian in the particle distribution function \cite{NumRec}. In other words, collisional effects become more and more efficient as kinetic effects work to perturb the particle distribution function and generate fine velocity scales \cite{pezzicoll}.

\subsection{Free energy terms evolution} \label{ener}
To complete our analysis, we focus here on the time evolution of the terms of the free energy conservation law discussed in Sec.~\ref{en_cons}. In particular, by monitoring the total budget in each simulation, one can point out the role of linear and nonlinear terms in the electric energy, the free energy stored in the distribution function and the contribution of collisions. Moreover, it is possible to investigate the limits of the approximation of Eq.~\eqref{budget_int_appr}, observing whether it is valid in a linear simulation at any time, but also quantifying in nonlinear runs the role of high-order terms in the Taylor expansion of Eq.~\eqref{budget_int}. For this purpose, one can directly compare $\mathcal{F}_1$ with $\mathcal{F}$ and $\mathcal{C}_1$ with $\mathcal{C}$, or write
\begin{align}
\mathcal{E} + \mathcal{F} - \mathcal{C} & = \mathcal{E} + \mathcal{F}_1 - \mathcal{C}_1 + \sum_{\beta = 2}^{\infty} \left( \mathcal{F}_\beta - \mathcal{C}_\beta \right) \nonumber \\
& = \mathcal{E} + \mathcal{F}_1 - \mathcal{C}_1 + \mathcal{F}_{>1} - \mathcal{C}_{>1} = C_0 
\end{align}
and evaluate the contribution of $\mathcal{F}_{>1} - \mathcal{C}_{>1}$ in the expression of $C_0$.

\begin{figure*}
  \begin{center}
        \includegraphics[scale=0.96]{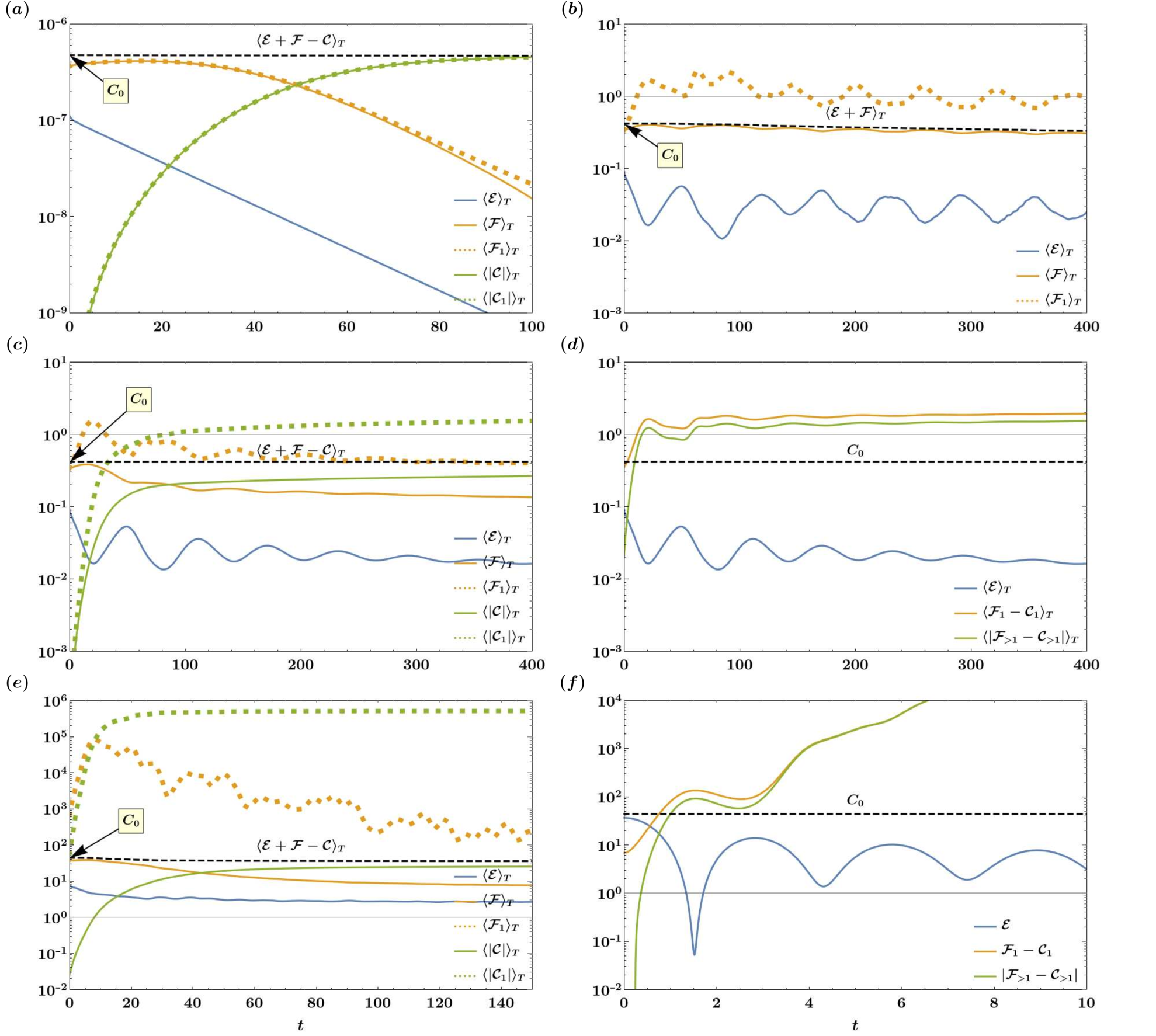} 
        \caption{Time evolution of terms of the free energy conservation law for Run II (collisional linear damping) [panel (a)], Run III (collisionless nonlinear trapping) [panel (b)], Run IV (collisional nonlinear trapping) [panels (c)-(d)], and Run VII (Langmuir turbulence) [panels (e)-(f)]. Panels (a)-(c) and (e): comparison of $\mathcal{E}$, $\mathcal{F}$ and $\mathcal{C}$ (solid lines) with the first-order approximations $\mathcal{E}_1$ and $\mathcal{F}_1$ (dotted lines) and the sum $\mathcal{E} + \mathcal{F} - \mathcal{C}$ (dashed black line), which remains almost equal to the initial budget $C_0$ in each simulation. Panels (d) and (f): comparison of $\mathcal{E}$, $\mathcal{F}_1-\mathcal{C}_1$ and $\mathcal{F}_{>1}-\mathcal{C}_{>1}$ (solid lines) with $C_0$ (dashed black line). Dashed blue line in panel (a): Landau damping trend $e^{2 \gamma t}$ estimated in Run I ($\gamma \simeq -2.57 \times 10^{-2}$). $\langle ... \rangle_T$ denotes the average over the period of Langmuir waves $T$.}
        \label{fig8}
  \end{center}
\end{figure*}

The results of this analysis are reported in Fig.~\ref{fig8} for Run II [panel (a)], Run III [panel (b)], Run IV [panels (c)-(d)], and Run VII [panels (e)-(f)]. In panels (a)-(e) of this figure, the oscillations at the Langmuir frequency have been smoothed out, by averaging the signals over $T$, i.e. the typical oscillation period of the Langmuir fluctuations. 

For the case of linear damping, in the absence of collisions (Run I, not shown here) we observe the exponential decay of $\mathcal{E}$ due to Landau damping: the electric energy loss is balanced by the growth of the free energy associated with $f$, which triggers the filamentation process. The total budget fluctuates within a range of $\simeq 0.2 \%$ of $C_0$ during the whole simulation. As shown in Fig.~\ref{fig8} (a), when collisions are considered, $\langle\left| \mathcal{C} \right|\rangle_T$ (solid green curve) increases very rapidly and the dissipated energy soon becomes dominant; $\langle \mathcal{F} \rangle_T$ (solid orange curve) starts decaying after $t = 20$, as the distribution is driven by collisions towards thermal equilibrium. $\langle \left| \mathcal{C}_1 \right| \rangle_T$ and $\langle\mathcal{F}_1 \rangle_T$ in function of $t$ are indicated by green and orange dots. The fact that green and orange dots fall with good approximation on the orange and green solid curves for almost the entire time evolution demonstrates that the contribution of the nonlinear terms $\mathcal{F}_{>1}$ and $\mathcal{C}_{>1}$ remains negligible. We also observe that the trend of $\mathcal{E}$ (solid blue curve) is very similar to the decay rate $\gamma \simeq -2.57 \times 10^{-2}$ expected for the collisionless case (dashed blue line). The total free energy (dashed black line) slightly decreases in time, with a limited loss of $\simeq 4 \%$ at $t = T_{max}$.

In Run III (the collisionless simulation of nonlinear trapping), represented in Fig.~\ref{fig8} (b), $\langle \mathcal{E} \rangle_T$ (solid blue curve) displays typical trapping oscillations. $\langle \mathcal{F} \rangle_T$ (solid orange curve) oscillates correspondingly with an opposite phase, such that their sum remains almost constant. Interestingly, in this case, $\langle \mathcal{F} \rangle_T$ departs from the linear contribution $\langle \mathcal{F}_1 \rangle_T$ (orange dots), as expected when the system dynamics is dominated by nonlinear effects.

In Run IV [Fig.~\ref{fig8} (c)], collisions are turned on in the case of nonlinear particle trapping and dominate (solid green curve) the time evolution of the free energy, already after one trapping oscillation. As a consequence, $\langle \mathcal{F}_1 \rangle_T$ (solid orange curve) decreases as collisions work to smooth out fine velocity gradients. Accordingly, the electric energy (solid blue curve) undergoes trapping oscillations, damped in time due to collisions. It is important to point out that $\langle \mathcal{F}_1 \rangle_T$ (orange dots) and $\langle \left| \mathcal{C}_1 \right| \rangle_T$ (green dots) display a qualitatively similar behavior as $\langle \mathcal{F} \rangle_T$ and $\langle \left| \mathcal{C} \right| \rangle_T$, but significant quantitative differences are evident, this meaning that nonlinear effects are dominant. The evolution of the total budget is shown as usual as a dashed black line. For the same Run IV, in Fig.~\ref{fig8} (d), together with $\langle \mathcal{E} \rangle_T$, we report the time evolution of $\langle \mathcal{F}_1 - \mathcal{C}_1 \rangle_T$ (orange curve) and $\langle \left| \mathcal{F}_{>1} - \mathcal{C}_{>1} \right| \rangle_T$ (green curve). This panel shows even more clearly that terms with $\beta > 1$ increase of almost two orders of magnitude in a time interval of length $\tau$, so the approximation of Eq.~\eqref{budget_int_appr} is invalid.

The same analysis has been performed for Run VII in panels (e)-(f) of Fig.~\ref{fig8}. The conclusions are qualitatively similar to those for Run IV, but the turbulent character of the system evolution in Run VII makes the role of nonlinear effects even more evident. In particular, we notice that $\langle \left| \mathcal{C} \right| \rangle_T$ (solid green curve), becomes dominant in a shorter time than in Run IV ($t \simeq 40$ rather than $t \simeq 100$): after that, the electric energy, the free energy of the distribution function and the collisional term keep the same order of magnitude. At the same time, the deviation of $\langle \mathcal{F}_1$ (orange dots) and $\langle \left| \mathcal{C}_1 \right| \rangle_T$ (green dots) from their exact values is even stronger than in the nonlinear trapping case, so $\beta > 1$ terms are necessary to conserve the total budget (dashed black line). The growth rate of these terms can be appreciated in Fig.~\ref{fig8} (f), which shows how $ \left| \mathcal{F}_{>1} - \mathcal{C}_{>1} \right|$ (green curve) reaches the same order of magnitude of $\mathcal{F}_{1} - \mathcal{C}_{1}$ (orange curve) in a time interval comparable to the trapping time of the highest Fourier harmonic excited at the beginning of the simulation, i.e. $\tau (k_{10}) \simeq{1.9}$. These examples may be important for space and laboratory plasma turbulence, where the amount of perturbations is significant.

\section{Conclusions} \label{secIV}
In this paper, we have solved the 1D-1V Vlasov-Poisson system assuming fixed ions, simulating both collisional and collisionless plasma regimes. We have performed a full spectral decomposition of the electron distribution function from the simulations, by employing the discrete Fourier-Hermite transform. This has allowed us to efficiently analyze the nature of free energy cascades towards small scales both in physical and in velocity space, and how these are influenced by the effect of collisions.

At first, we have analyzed the enstrophy flow in the velocity space through the Hermite transform. During the time evolution in linear regimes, we have observed that enstrophy is highly peaked on specific Hermite modes, consistently with a filamentation process expected within the Landau scenario. Then, we have focused on nonlinear regimes and we have noticed that cascades are particularly active and quickly involve all available modes, especially in the cases of instabilities and turbulence. In particular, after an initial phase when linear filamentation and nonlinear energy spread coexist, the latter tends to get dominant, resulting in spectra which follow power laws in agreement with previous expectations. As time goes on, the collisional term tends to suppress the highest modes and eventually it appears to have a strong influence on the whole spectrum. In the fully turbulent case, which can be relevant for space and highly turbulent laboratory plasmas, the net transfer among physical and velocity space is very large, and the picture is similar to a Kolmogorov-like cascade \cite{MMS}.

Then we have completed the analysis by performing the Fourier transform in physical space: here, we observe an analogous cascade towards small scales which seems highly dependent on the initial conditions, but less efficient than the Hermite cascade in all nonlinear runs as it cannot reach the highest available Fourier modes, resulting in anisotropic Fourier-Hermite spectra. Even the physical cascade is strongly damped by collisions, as one can notice from the inspection of the time evolution of Fourier-Hermite spectra of the distribution function and of the Fourier spectra of the self-induced electric field.

We have also obtained a conservation law which quantifies the dissipation rate of the electric energy and the free energy associated with the distribution function due to the collisional term. This law has been Taylor expanded by using a power series of the perturbation of the distribution function with respect to Maxwell-Boltzmann equilibrium, allowing to compare the role of linear and nonlinear contributions. We conclude that our code respects the total free energy budget with a relatively small error and that, in nonlinear runs, the linear approximation of the conservation law, obtained by truncating the Taylor expansion, deviates from the exact law of several orders of magnitude. In fact, the high-order terms of the Taylor expansion become relevant at time scales of the order of the trapping time, so they can be neglected in linear runs only.

The findings of this research demonstrate the effectiveness of the application of our FHT algorithm in the analysis of particle distribution functions and the possibility of distinguishing one regime from another by determining the power law of the corresponding Fourier-Hermite spectra. The versatility of our method permits an easy extension of our FHT code to the more complex and realistic 2D and 3D plasma simulations, including the magnetic field, with the possibility to directly investigate data collected by new generations of space probes, such as new concepts of spacecraft constellations.

\begin{acknowledgments}

The simulations have been performed at the Newton cluster at University of Calabria and the work is supported by ‘Progetto STAR 2-PIR01 00008’ (Italian Ministry of University and Research). The authors acknowledge supercomputing resources and support from ICSC - Centro Nazionale di Ricerca in High Performance Computing, Big Data and Quantum Computing - and hosting entity, funded by European Union - NextGenerationEU.

\end{acknowledgments}

\section*{Author declarations}

\subsection*{Conflict of interest}

The authors have no conflicts to disclose.

\section*{Data availability}

The data that support the findings of this study are available from the corresponding author upon reasonable request.

\appendix

\section{Gauss-Hermite functions asymptotic behavior} \label{AppA}
For large values of the index $m$, the $m$-th Gauss-Hermite function $\psi_m(v)$ can be approximated as:
\begin{equation} \label{approx1}
\psi_m(v) \sim C_m \cos \left( \sqrt{2m} v - \frac{m \pi}{2} \right) .  
\end{equation}
Eq.~(\ref{approx1}) derives directly from the following asymptotic behavior of Hermite polynomials (see Ref.~\onlinecite{AbrSte1}):
\begin{equation} \label{approx2}
H_m(v)e^{-\frac{v^2}{2}} \sim \frac{2\Gamma(m)}{\Gamma(\frac{m}{2})} \cos \left( \sqrt{2m} v - \frac{m \pi}{2} \right) ,
\end{equation}
where $\Gamma$ denotes the gamma function.

As discussed in Sec.~\ref{Her}, even without providing the explicit form of $C_m$, Eq.~\eqref{approx1} allows to understand the time evolution of $m^\star(t)$ when linear filamentation is at work. However, to explain the trend of the power law $m^{-1 / 2}$ of Fig.~\ref{fig5} (a), it is necessary to find the asymptotic trend of $C_m$. From the definition of $\psi_m$ and \eqref{approx2}:
\begin{equation}
C_m \sim \frac{1}{\sqrt{2^m m! \sqrt{\pi}}} \frac{2 \Gamma(m)}{\Gamma(\frac{m}{2})} .
\end{equation}
Thanks to the application of Stirling's formula \cite{AbrSte2}:
\begin{subequations}
\begin{gather}
z! \sim \sqrt{2 \pi z} \left( \frac{z}{e} \right)^z , \\[5pt]
\Gamma(z) \sim \sqrt{\frac{2 \pi}{z}} \left( \frac{z}{e} \right)^z ,
\end{gather}
\end{subequations}
it is possible to deduce that $C_m \sim \sqrt[4]{2 / \left( m \pi^2  \right)}$. Using this simplified expression for $C_m$, and considering only even modes because of the parity of $f$ in Run I, one has:
\usetagform{nowidth}
\begin{widetext}
\begin{align}
f_{m^\star} (x) =& \int_{-\infty}^{+\infty} \delta f\left( x,v,\frac{\sqrt{2m^\star}}{k_0} \right) \psi_{m^\star}(v) dv \sim (-1)^{\frac{m^\star}{2}} \frac{1}{\sqrt{\pi}} \sqrt[4]{\frac{2}{m}} \frac{A}{\sqrt{2 \pi}} \int_{-\infty}^{+\infty} \cos \left( \sqrt{2m^\star} v  \right) \cos \left( k_0 x - \sqrt{2m^\star} v  \right) e^{-\frac{v^2}{2}} dv \nonumber \\
=& (-1)^{\frac{m^\star}{2}} \frac{A}{\sqrt{\pi \sqrt{8 m^\star}}} \left( 1 + e^{-4 m^\star} \right) \cos (k_0 x) \sim (-1)^{\frac{m^\star}{2}} \frac{A}{\sqrt{\pi \sqrt{8 m^\star}}} \cos (k_0 x) \Rightarrow \langle f_{m^\star}^2 \rangle_x \propto \frac{1}{\sqrt{m^\star}} ,
\end{align}
\end{widetext}
\usetagform{default}
in agreement with the observed power law.
 
\section{Free energy conservation law derivation} \label{AppB}
The conservation equation \eqref{ECL} and its integrated form \eqref{budget_int} can be obtained directly from the dimensionless collisional Vlasov equation. By multiplying the latter by $v^2 / 2$ and integrating over the whole phase space, one gets:
\begin{equation}
\int \frac{v^2}{2} \left( \frac{\partial f}{\partial t} + v \frac{\partial f}{\partial x} - E \frac{\partial f}{\partial v} \right) dxdv = \int \frac{v^2}{2} \frac{\partial f}{\partial t} \bigr\rvert_{coll} dxdv . \label{cons}
\end{equation}
On the left side, the first term is equal to $(d / dt) \int (u + n_e U_e^2 / 2) dx$, where $u$ is the internal energy density of particles scaled by $n_0 m_e v_{th , e}^2$. The second term is null because of spatial periodicity. The third one gets the form (in 1D) $-\int JE dx$, where $J$ is the electron current density scaled by $n_0 e v_{th , e}$. By applying the Poynting theorem, one can see that $\int JE dx = - \left( d / dt \right) \int \left( E^2 / 2 \right) dx$, so the left side of \eqref{cons} is the time derivative of total energy density; moreover, the Dougherty operator conserves energy and the right-hand side of \eqref{cons} is null:
\begin{equation} \label{energia}
\frac{d}{dt} \int \left( u + \frac{n_e U_e^2}{2} + \frac{E^2}{2} \right) dx = 0 .
\end{equation}
According to Boltzmann's H-theorem \cite{Boltzmann}, the time derivative of entropy $S$ (scaled by $n_0 k_B \lambda_{D , e}$) is:
\begin{align}
\frac{dS}{dt} =& - \frac{d}{dt} \int f \ln f dxdv = - \int \ln f \frac{\partial f}{\partial t} dxdv \nonumber \\ 
=& - \int \ln f \left( \frac{\partial f}{\partial t} \bigr\rvert_{coll} - v \frac{\partial f}{\partial x} + E \frac{\partial f}{\partial v} \right) dxdv , \label{entr_1}
\end{align}
where the last two terms of the right-hand side are null, due to boundary conditions of $f$ in $x$ and $v$. Eq.~\eqref{entr_1} can be manipulated as follows:
\begin{align} 
\frac{dS}{dt} =& - \int \left[ \ln f_{MB} + \ln \left( \frac{f}{f_{MB}} \right) \right] \frac{\partial f}{\partial t} \bigr\rvert_{coll} dxdv \nonumber \\
=& - \int \left[ \ln \left( \frac{1}{\sqrt{2 \pi}} \right) - \frac{v^2}{2} \right] \frac{\partial f}{\partial t} \bigr\rvert_{coll} dxdv  \nonumber \\
& - \int \ln \left( \frac{f}{f_{MB}} \right) \frac{\partial f}{\partial t} \bigr\rvert_{coll} dxdv . \label{entr_2}
\end{align}
The first integral in the previous equation vanishes, as $\int \partial f / \partial t \rvert_{coll} dxdv = 0$  and $\int v^2 \partial f / \partial t \rvert_{coll} dxdv = 0$, in order to guarantee the conservation of the number of particles and the total energy (as in Eq.~\eqref{cons}).

At the same time, one can write:
\begin{align}
\frac{dS}{dt} =& - \frac{d}{dt} \int f \left[ \ln f_{MB} + \ln \left( \frac{f}{f_{MB}} \right) \right] dxdv \nonumber \\
=& - \ln \left( \frac{1}{\sqrt{2 \pi}} \right) \frac{d}{dt} \int f dx dv + \frac{d}{dt} \int \frac{v^2}{2} f dx dv \nonumber \\
& - \frac{d}{dt} \int f \ln \left( \frac{f}{f_{MB}} \right) dxdv. \label{entr_3}
\end{align}
The first term on the right side is null due to mass conservation, while the second is $(d / dt) \int (u + n_e U_e^2 / 2) dx$. From Eqs.~\eqref{energia}, \eqref{entr_2} and \eqref{entr_3}, we get Eq.~\eqref{ECL}. Note that Eq.~\eqref{energia} is already an exact law able to describe the energy flow from the electric field to the system of particles and vice versa, but it cannot effectively describe the role of the collisional term. On the other hand, the inclusion of entropy in the conservation law, like in Eq.~\eqref{ECL}, permits highlighting the influence of collisions on the dynamics of the system.

When $\Delta f = f - f_{MB} \ll f_{MB}$, it is possible to Taylor expand $\ln \left( f / f_{MB} \right)$ and $f \ln \left( f / f_{MB} \right)$. Taking into account that $\int \Delta f dxdv = 0$, one finds:
\begin{equation}
\frac{d}{dt} \left( \int \frac{E^2}{2} dx + \int \frac{\Delta f^2}{2 f_{MB}} dxdv \right) \approx \int \frac{\Delta f}{f_{MB}} \frac{\partial f}{\partial t} \bigr\rvert_{coll} dxdv , 
\end{equation}
which corresponds to the free energy conservation law in Refs.~\onlinecite{MixAdv, Budget} and, once integrated over $t$, becomes Eq.~\eqref{budget_int_appr}.

\nocite{*}
\bibliography{aipsamp}

\providecommand{\noopsort}[1]{}\providecommand{\singleletter}[1]{#1}%
\begin{thebibliography}{55}%
\makeatletter
\providecommand \@ifxundefined [1]{%
 \@ifx{#1\undefined}
}%
\providecommand \@ifnum [1]{%
 \ifnum #1\expandafter \@firstoftwo
 \else \expandafter \@secondoftwo
 \fi
}%
\providecommand \@ifx [1]{%
 \ifx #1\expandafter \@firstoftwo
 \else \expandafter \@secondoftwo
 \fi
}%
\providecommand \natexlab [1]{#1}%
\providecommand \enquote  [1]{``#1''}%
\providecommand \bibnamefont  [1]{#1}%
\providecommand \bibfnamefont [1]{#1}%
\providecommand \citenamefont [1]{#1}%
\providecommand \href@noop [0]{\@secondoftwo}%
\providecommand \href [0]{\begingroup \@sanitize@url \@href}%
\providecommand \@href[1]{\@@startlink{#1}\@@href}%
\providecommand \@@href[1]{\endgroup#1\@@endlink}%
\providecommand \@sanitize@url [0]{\catcode `\\12\catcode `\$12\catcode `\&12\catcode `\#12\catcode `\^12\catcode `\_12\catcode `\%12\relax}%
\providecommand \@@startlink[1]{}%
\providecommand \@@endlink[0]{}%
\providecommand \url  [0]{\begingroup\@sanitize@url \@url }%
\providecommand \@url [1]{\endgroup\@href {#1}{\urlprefix }}%
\providecommand \urlprefix  [0]{URL }%
\providecommand \Eprint [0]{\href }%
\providecommand \doibase [0]{http://dx.doi.org/}%
\providecommand \selectlanguage [0]{\@gobble}%
\providecommand \bibinfo  [0]{\@secondoftwo}%
\providecommand \bibfield  [0]{\@secondoftwo}%
\providecommand \translation [1]{[#1]}%
\providecommand \BibitemOpen [0]{}%
\providecommand \bibitemStop [0]{}%
\providecommand \bibitemNoStop [0]{.\EOS\space}%
\providecommand \EOS [0]{\spacefactor3000\relax}%
\providecommand \BibitemShut  [1]{\csname bibitem#1\endcsname}%
\let\auto@bib@innerbib\@empty
\bibitem [{\citenamefont {Servidio}\ \emph {et~al.}(2017)\citenamefont {Servidio}, \citenamefont {Chasapis}, \citenamefont {Matthaeus}, \citenamefont {Perrone}, \citenamefont {Valentini}, \citenamefont {Parashar}, \citenamefont {Veltri}, \citenamefont {Gershman}, \citenamefont {Russell}, \citenamefont {Giles}, \citenamefont {Fuselier}, \citenamefont {Phan},\ and\ \citenamefont {Burch}}]{MMS}%
  \BibitemOpen
  \bibfield  {author} {\bibinfo {author} {\bibfnamefont {S.}~\bibnamefont {Servidio}}, \bibinfo {author} {\bibfnamefont {A.}~\bibnamefont {Chasapis}}, \bibinfo {author} {\bibfnamefont {W.~H.}\ \bibnamefont {Matthaeus}}, \bibinfo {author} {\bibfnamefont {D.}~\bibnamefont {Perrone}}, \bibinfo {author} {\bibfnamefont {F.}~\bibnamefont {Valentini}}, \bibinfo {author} {\bibfnamefont {T.~N.}\ \bibnamefont {Parashar}}, \bibinfo {author} {\bibfnamefont {P.}~\bibnamefont {Veltri}}, \bibinfo {author} {\bibfnamefont {D.}~\bibnamefont {Gershman}}, \bibinfo {author} {\bibfnamefont {C.~T.}\ \bibnamefont {Russell}}, \bibinfo {author} {\bibfnamefont {B.}~\bibnamefont {Giles}}, \bibinfo {author} {\bibfnamefont {S.~A.}\ \bibnamefont {Fuselier}}, \bibinfo {author} {\bibfnamefont {T.~D.}\ \bibnamefont {Phan}}, \ and\ \bibinfo {author} {\bibfnamefont {J.}~\bibnamefont {Burch}},\ }\href@noop {} {\bibfield  {journal} {\bibinfo  {journal} {Phys. Rev. Lett.}\ }\textbf {\bibinfo {volume} {119\normalfont{(20)}}},\ \bibinfo {pages}
  {205101} (\bibinfo {year} {2017})}\BibitemShut {NoStop}%
\bibitem [{\citenamefont {Klein}\ and\ \citenamefont {Howes}(2016)}]{Klein}%
  \BibitemOpen
  \bibfield  {author} {\bibinfo {author} {\bibfnamefont {K.~G.}\ \bibnamefont {Klein}}\ and\ \bibinfo {author} {\bibfnamefont {G.~G.}\ \bibnamefont {Howes}},\ }\href@noop {} {\bibfield  {journal} {\bibinfo  {journal} {Astrophys. J. Lett.}\ }\textbf {\bibinfo {volume} {826\normalfont{(2)}}},\ \bibinfo {pages} {L30} (\bibinfo {year} {2016})}\BibitemShut {NoStop}%
\bibitem [{\citenamefont {Krall}\ and\ \citenamefont {Trivelpiece}(1973{\natexlab{a}})}]{VlaMax}%
  \BibitemOpen
  \bibfield  {author} {\bibinfo {author} {\bibfnamefont {N.~A.}\ \bibnamefont {Krall}}\ and\ \bibinfo {author} {\bibfnamefont {A.~W.}\ \bibnamefont {Trivelpiece}},\ }\href@noop {} {\emph {\bibinfo {title} {Principles of plasma physics}}}\ (\bibinfo  {publisher} {McGraw-Hill},\ \bibinfo {address} {New York},\ \bibinfo {year} {1973})\ pp.\ \bibinfo {pages} {369--375}\BibitemShut {NoStop}%
\bibitem [{\citenamefont {Schumer}\ and\ \citenamefont {Holloway}(1998)}]{FH1}%
  \BibitemOpen
  \bibfield  {author} {\bibinfo {author} {\bibfnamefont {J.~W.}\ \bibnamefont {Schumer}}\ and\ \bibinfo {author} {\bibfnamefont {J.~P.}\ \bibnamefont {Holloway}},\ }\href@noop {} {\bibfield  {journal} {\bibinfo  {journal} {J. Comput. Phys}\ }\textbf {\bibinfo {volume} {144\normalfont{(2)}}},\ \bibinfo {pages} {626} (\bibinfo {year} {1998})}\BibitemShut {NoStop}%
\bibitem [{\citenamefont {Parker}\ and\ \citenamefont {Dellar}(2015)}]{FH2}%
  \BibitemOpen
  \bibfield  {author} {\bibinfo {author} {\bibfnamefont {J.~T.}\ \bibnamefont {Parker}}\ and\ \bibinfo {author} {\bibfnamefont {P.~J.}\ \bibnamefont {Dellar}},\ }\href@noop {} {\bibfield  {journal} {\bibinfo  {journal} {J. Plasma Phys.}\ }\textbf {\bibinfo {volume} {81\normalfont{(2)}}},\ \bibinfo {pages} {305810203} (\bibinfo {year} {2015})}\BibitemShut {NoStop}%
\bibitem [{\citenamefont {Delzanno}(2015)}]{FH3}%
  \BibitemOpen
  \bibfield  {author} {\bibinfo {author} {\bibfnamefont {G.~L.}\ \bibnamefont {Delzanno}},\ }\href@noop {} {\bibfield  {journal} {\bibinfo  {journal} {J. Comput. Phys.}\ }\textbf {\bibinfo {volume} {301}},\ \bibinfo {pages} {338} (\bibinfo {year} {2015})}\BibitemShut {NoStop}%
\bibitem [{\citenamefont {Loureiro}\ \emph {et~al.}(2016)\citenamefont {Loureiro}, \citenamefont {Dorland}, \citenamefont {Fazendeiro}, \citenamefont {Kanekar}, \citenamefont {Mallet}, \citenamefont {Vilelas},\ and\ \citenamefont {Zocco}}]{FH4}%
  \BibitemOpen
  \bibfield  {author} {\bibinfo {author} {\bibfnamefont {N.~F.}\ \bibnamefont {Loureiro}}, \bibinfo {author} {\bibfnamefont {W.}~\bibnamefont {Dorland}}, \bibinfo {author} {\bibfnamefont {L.}~\bibnamefont {Fazendeiro}}, \bibinfo {author} {\bibfnamefont {A.}~\bibnamefont {Kanekar}}, \bibinfo {author} {\bibfnamefont {A.}~\bibnamefont {Mallet}}, \bibinfo {author} {\bibfnamefont {M.~S.}\ \bibnamefont {Vilelas}}, \ and\ \bibinfo {author} {\bibfnamefont {A.}~\bibnamefont {Zocco}},\ }\href@noop {} {\bibfield  {journal} {\bibinfo  {journal} {Comput. Phys. Commun.}\ }\textbf {\bibinfo {volume} {206}},\ \bibinfo {pages} {45} (\bibinfo {year} {2016})}\BibitemShut {NoStop}%
\bibitem [{\citenamefont {Roytershteyn}\ and\ \citenamefont {Delzanno}(2018)}]{FH5}%
  \BibitemOpen
  \bibfield  {author} {\bibinfo {author} {\bibfnamefont {V.}~\bibnamefont {Roytershteyn}}\ and\ \bibinfo {author} {\bibfnamefont {G.~L.}\ \bibnamefont {Delzanno}},\ }\href@noop {} {\bibfield  {journal} {\bibinfo  {journal} {Front. Astron. Space Sci.}\ }\textbf {\bibinfo {volume} {5}},\ \bibinfo {pages} {27} (\bibinfo {year} {2018})}\BibitemShut {NoStop}%
\bibitem [{\citenamefont {Valentini}\ \emph {et~al.}(2005{\natexlab{a}})\citenamefont {Valentini}, \citenamefont {Carbone}, \citenamefont {Veltri},\ and\ \citenamefont {Mangeney}}]{VlaPoi}%
  \BibitemOpen
  \bibfield  {author} {\bibinfo {author} {\bibfnamefont {F.}~\bibnamefont {Valentini}}, \bibinfo {author} {\bibfnamefont {V.}~\bibnamefont {Carbone}}, \bibinfo {author} {\bibfnamefont {P.}~\bibnamefont {Veltri}}, \ and\ \bibinfo {author} {\bibfnamefont {A.}~\bibnamefont {Mangeney}},\ }\href@noop {} {\bibfield  {journal} {\bibinfo  {journal} {Phys. Rev. E}\ }\textbf {\bibinfo {volume} {71\normalfont{(1)}}},\ \bibinfo {pages} {017402} (\bibinfo {year} {2005}{\natexlab{a}})}\BibitemShut {NoStop}%
\bibitem [{\citenamefont {Pezzi}\ \emph {et~al.}(2013)\citenamefont {Pezzi}, \citenamefont {Valentini}, \citenamefont {Perrone},\ and\ \citenamefont {Veltri}}]{EulColl}%
  \BibitemOpen
  \bibfield  {author} {\bibinfo {author} {\bibfnamefont {O.}~\bibnamefont {Pezzi}}, \bibinfo {author} {\bibfnamefont {F.}~\bibnamefont {Valentini}}, \bibinfo {author} {\bibfnamefont {D.}~\bibnamefont {Perrone}}, \ and\ \bibinfo {author} {\bibfnamefont {P.}~\bibnamefont {Veltri}},\ }\href@noop {} {\bibfield  {journal} {\bibinfo  {journal} {Phys. Plasmas}\ }\textbf {\bibinfo {volume} {20\normalfont{(9)}}},\ \bibinfo {pages} {092111} (\bibinfo {year} {2013})}\BibitemShut {NoStop}%
\bibitem [{\citenamefont {Pezzi}\ \emph {et~al.}(2016{\natexlab{a}})\citenamefont {Pezzi}, \citenamefont {Camporeale},\ and\ \citenamefont {Valentini}}]{NumRec}%
  \BibitemOpen
  \bibfield  {author} {\bibinfo {author} {\bibfnamefont {O.}~\bibnamefont {Pezzi}}, \bibinfo {author} {\bibfnamefont {E.}~\bibnamefont {Camporeale}}, \ and\ \bibinfo {author} {\bibfnamefont {F.}~\bibnamefont {Valentini}},\ }\href@noop {} {\bibfield  {journal} {\bibinfo  {journal} {Phys. Plasmas}\ }\textbf {\bibinfo {volume} {23\normalfont{(2)}}},\ \bibinfo {pages} {022103} (\bibinfo {year} {2016}{\natexlab{a}})}\BibitemShut {NoStop}%
\bibitem [{\citenamefont {Dougherty}(1964)}]{Dough1}%
  \BibitemOpen
  \bibfield  {author} {\bibinfo {author} {\bibfnamefont {J.~P.}\ \bibnamefont {Dougherty}},\ }\href@noop {} {\bibfield  {journal} {\bibinfo  {journal} {Phys. Fluids}\ }\textbf {\bibinfo {volume} {7\normalfont{(11)}}},\ \bibinfo {pages} {1788} (\bibinfo {year} {1964})}\BibitemShut {NoStop}%
\bibitem [{\citenamefont {Dougherty}\ and\ \citenamefont {Watson}(1967)}]{Dough2}%
  \BibitemOpen
  \bibfield  {author} {\bibinfo {author} {\bibfnamefont {J.~P.}\ \bibnamefont {Dougherty}}\ and\ \bibinfo {author} {\bibfnamefont {S.~R.}\ \bibnamefont {Watson}},\ }\href@noop {} {\bibfield  {journal} {\bibinfo  {journal} {J. Plasma Phys.}\ }\textbf {\bibinfo {volume} {1}},\ \bibinfo {pages} {317} (\bibinfo {year} {1967})}\BibitemShut {NoStop}%
\bibitem [{\citenamefont {Krall}\ and\ \citenamefont {Trivelpiece}(1973{\natexlab{b}})}]{bump}%
  \BibitemOpen
  \bibfield  {author} {\bibinfo {author} {\bibfnamefont {N.~A.}\ \bibnamefont {Krall}}\ and\ \bibinfo {author} {\bibfnamefont {A.~W.}\ \bibnamefont {Trivelpiece}},\ }\href@noop {} {\emph {\bibinfo {title} {Principles of plasma physics}}}\ (\bibinfo  {publisher} {McGraw-Hill},\ \bibinfo {address} {New York},\ \bibinfo {year} {1973})\ pp.\ \bibinfo {pages} {458--463}\BibitemShut {NoStop}%
\bibitem [{\citenamefont {Krall}\ and\ \citenamefont {Trivelpiece}(1973{\natexlab{c}})}]{beam}%
  \BibitemOpen
  \bibfield  {author} {\bibinfo {author} {\bibfnamefont {N.~A.}\ \bibnamefont {Krall}}\ and\ \bibinfo {author} {\bibfnamefont {A.~W.}\ \bibnamefont {Trivelpiece}},\ }\href@noop {} {\emph {\bibinfo {title} {Principles of plasma physics}}}\ (\bibinfo  {publisher} {McGraw-Hill},\ \bibinfo {address} {New York},\ \bibinfo {year} {1973})\ pp.\ \bibinfo {pages} {449--458}\BibitemShut {NoStop}%
\bibitem [{\citenamefont {Landau}(1946)}]{Landau}%
  \BibitemOpen
  \bibfield  {author} {\bibinfo {author} {\bibfnamefont {L.~D.}\ \bibnamefont {Landau}},\ }\href@noop {} {\bibfield  {journal} {\bibinfo  {journal} {J. Phys. (Moscow)}\ }\textbf {\bibinfo {volume} {10}},\ \bibinfo {pages} {25} (\bibinfo {year} {1946})}\BibitemShut {NoStop}%
\bibitem [{\citenamefont {O'Neil}(1965)}]{ONeil}%
  \BibitemOpen
  \bibfield  {author} {\bibinfo {author} {\bibfnamefont {T.}~\bibnamefont {O'Neil}},\ }\href@noop {} {\bibfield  {journal} {\bibinfo  {journal} {Phys. Fluids}\ }\textbf {\bibinfo {volume} {8\normalfont{(12)}}},\ \bibinfo {pages} {2255} (\bibinfo {year} {1965})}\BibitemShut {NoStop}%
\bibitem [{\citenamefont {Pezzi}\ \emph {et~al.}(2019{\natexlab{a}})\citenamefont {Pezzi}, \citenamefont {Valentini}, \citenamefont {Servidio}, \citenamefont {Camporeale},\ and\ \citenamefont {Veltri}}]{FouHer}%
  \BibitemOpen
  \bibfield  {author} {\bibinfo {author} {\bibfnamefont {O.}~\bibnamefont {Pezzi}}, \bibinfo {author} {\bibfnamefont {F.}~\bibnamefont {Valentini}}, \bibinfo {author} {\bibfnamefont {S.}~\bibnamefont {Servidio}}, \bibinfo {author} {\bibfnamefont {E.}~\bibnamefont {Camporeale}}, \ and\ \bibinfo {author} {\bibfnamefont {P.}~\bibnamefont {Veltri}},\ }\href@noop {} {\bibfield  {journal} {\bibinfo  {journal} {Plasma Phys. Control. Fusion}\ }\textbf {\bibinfo {volume} {61\normalfont{(5)}}},\ \bibinfo {pages} {054005} (\bibinfo {year} {2019}{\natexlab{a}})}\BibitemShut {NoStop}%
\bibitem [{\citenamefont {Schekochihin}\ \emph {et~al.}(2008)\citenamefont {Schekochihin}, \citenamefont {Cowley}, \citenamefont {Dorland}, \citenamefont {Hammett}, \citenamefont {Howes}, \citenamefont {Plunk}, \citenamefont {Quataert},\ and\ \citenamefont {Tatsuno}}]{Budget}%
  \BibitemOpen
  \bibfield  {author} {\bibinfo {author} {\bibfnamefont {A.~A.}\ \bibnamefont {Schekochihin}}, \bibinfo {author} {\bibfnamefont {S.~C.}\ \bibnamefont {Cowley}}, \bibinfo {author} {\bibfnamefont {W.}~\bibnamefont {Dorland}}, \bibinfo {author} {\bibfnamefont {G.~W.}\ \bibnamefont {Hammett}}, \bibinfo {author} {\bibfnamefont {G.~G.}\ \bibnamefont {Howes}}, \bibinfo {author} {\bibfnamefont {G.~G.}\ \bibnamefont {Plunk}}, \bibinfo {author} {\bibfnamefont {E.}~\bibnamefont {Quataert}}, \ and\ \bibinfo {author} {\bibfnamefont {T.}~\bibnamefont {Tatsuno}},\ }\href@noop {} {\bibfield  {journal} {\bibinfo  {journal} {Plasma Phys. Control. Fusion}\ }\textbf {\bibinfo {volume} {50\normalfont{(12)}}},\ \bibinfo {pages} {124024} (\bibinfo {year} {2008})}\BibitemShut {NoStop}%
\bibitem [{\citenamefont {Schekochihin}\ \emph {et~al.}(2016)\citenamefont {Schekochihin}, \citenamefont {Parker}, \citenamefont {Highcock}, \citenamefont {Dellar}, \citenamefont {Dorland},\ and\ \citenamefont {Hammett}}]{MixAdv}%
  \BibitemOpen
  \bibfield  {author} {\bibinfo {author} {\bibfnamefont {A.~A.}\ \bibnamefont {Schekochihin}}, \bibinfo {author} {\bibfnamefont {J.~T.}\ \bibnamefont {Parker}}, \bibinfo {author} {\bibfnamefont {E.~G.}\ \bibnamefont {Highcock}}, \bibinfo {author} {\bibfnamefont {P.~J.}\ \bibnamefont {Dellar}}, \bibinfo {author} {\bibfnamefont {W.}~\bibnamefont {Dorland}}, \ and\ \bibinfo {author} {\bibfnamefont {G.~W.}\ \bibnamefont {Hammett}},\ }\href@noop {} {\bibfield  {journal} {\bibinfo  {journal} {J. Plasma Phys.}\ }\textbf {\bibinfo {volume} {82\normalfont{(2)}}},\ \bibinfo {pages} {905820212} (\bibinfo {year} {2016})}\BibitemShut {NoStop}%
\bibitem [{\citenamefont {Budaev}\ \emph {et~al.}(2015)\citenamefont {Budaev}, \citenamefont {Zelenyi},\ and\ \citenamefont {Savin}}]{inter1}%
  \BibitemOpen
  \bibfield  {author} {\bibinfo {author} {\bibfnamefont {V.~P.}\ \bibnamefont {Budaev}}, \bibinfo {author} {\bibfnamefont {L.~M.}\ \bibnamefont {Zelenyi}}, \ and\ \bibinfo {author} {\bibfnamefont {S.~P.}\ \bibnamefont {Savin}},\ }\href@noop {} {\bibfield  {journal} {\bibinfo  {journal} {J. Plasma Phys.}\ }\textbf {\bibinfo {volume} {81\normalfont{(6)}}},\ \bibinfo {pages} {395810602} (\bibinfo {year} {2015})}\BibitemShut {NoStop}%
\bibitem [{\citenamefont {Cerri}\ \emph {et~al.}(2018)\citenamefont {Cerri}, \citenamefont {Kunz},\ and\ \citenamefont {Califano}}]{inter2}%
  \BibitemOpen
  \bibfield  {author} {\bibinfo {author} {\bibfnamefont {S.~S.}\ \bibnamefont {Cerri}}, \bibinfo {author} {\bibfnamefont {M.~W.}\ \bibnamefont {Kunz}}, \ and\ \bibinfo {author} {\bibfnamefont {F.}~\bibnamefont {Califano}},\ }\href@noop {} {\bibfield  {journal} {\bibinfo  {journal} {Astrophys. J. Lett.}\ }\textbf {\bibinfo {volume} {856\normalfont{(1)}}},\ \bibinfo {pages} {L13} (\bibinfo {year} {2018})}\BibitemShut {NoStop}%
\bibitem [{\citenamefont {Tavassoli}\ \emph {et~al.}(2021)\citenamefont {Tavassoli}, \citenamefont {Shoucri}, \citenamefont {Smolyakov}, \citenamefont {Papahn~Zadeh},\ and\ \citenamefont {Spiteri}}]{ions}%
  \BibitemOpen
  \bibfield  {author} {\bibinfo {author} {\bibfnamefont {A.}~\bibnamefont {Tavassoli}}, \bibinfo {author} {\bibfnamefont {M.}~\bibnamefont {Shoucri}}, \bibinfo {author} {\bibfnamefont {A.}~\bibnamefont {Smolyakov}}, \bibinfo {author} {\bibfnamefont {M.}~\bibnamefont {Papahn~Zadeh}}, \ and\ \bibinfo {author} {\bibfnamefont {R.~J.}\ \bibnamefont {Spiteri}},\ }\href@noop {} {\bibfield  {journal} {\bibinfo  {journal} {Phys. Plasmas}\ }\textbf {\bibinfo {volume} {28\normalfont{(2)}}},\ \bibinfo {pages} {022307} (\bibinfo {year} {2021})}\BibitemShut {NoStop}%
\bibitem [{\citenamefont {Carril}\ \emph {et~al.}(2023)\citenamefont {Carril}, \citenamefont {Gidi}, \citenamefont {Navarro},\ and\ \citenamefont {Araneda}}]{BGK}%
  \BibitemOpen
  \bibfield  {author} {\bibinfo {author} {\bibfnamefont {H.~A.}\ \bibnamefont {Carril}}, \bibinfo {author} {\bibfnamefont {J.~A.}\ \bibnamefont {Gidi}}, \bibinfo {author} {\bibfnamefont {R.~E.}\ \bibnamefont {Navarro}}, \ and\ \bibinfo {author} {\bibfnamefont {J.~A.}\ \bibnamefont {Araneda}},\ }\href@noop {} {\bibfield  {journal} {\bibinfo  {journal} {Phys. Rev. E}\ }\textbf {\bibinfo {volume} {107\normalfont{(6)}}},\ \bibinfo {pages} {065203} (\bibinfo {year} {2023})}\BibitemShut {NoStop}%
\bibitem [{\citenamefont {Hakim}\ \emph {et~al.}(2020)\citenamefont {Hakim}, \citenamefont {Francisquez}, \citenamefont {Juno},\ and\ \citenamefont {Hammett}}]{Hakim}%
  \BibitemOpen
  \bibfield  {author} {\bibinfo {author} {\bibfnamefont {A.}~\bibnamefont {Hakim}}, \bibinfo {author} {\bibfnamefont {M.}~\bibnamefont {Francisquez}}, \bibinfo {author} {\bibfnamefont {J.}~\bibnamefont {Juno}}, \ and\ \bibinfo {author} {\bibfnamefont {G.~W.}\ \bibnamefont {Hammett}},\ }\href@noop {} {\bibfield  {journal} {\bibinfo  {journal} {J. Plasma Phys.}\ }\textbf {\bibinfo {volume} {86\normalfont{(4)}}},\ \bibinfo {pages} {905860403} (\bibinfo {year} {2020})}\BibitemShut {NoStop}%
\bibitem [{\citenamefont {Mangeney}\ \emph {et~al.}(2002)\citenamefont {Mangeney}, \citenamefont {Califano}, \citenamefont {Cavazzoni},\ and\ \citenamefont {Travnicek}}]{VanLeer}%
  \BibitemOpen
  \bibfield  {author} {\bibinfo {author} {\bibfnamefont {A.}~\bibnamefont {Mangeney}}, \bibinfo {author} {\bibfnamefont {F.}~\bibnamefont {Califano}}, \bibinfo {author} {\bibfnamefont {C.}~\bibnamefont {Cavazzoni}}, \ and\ \bibinfo {author} {\bibfnamefont {P.}~\bibnamefont {Travnicek}},\ }\href@noop {} {\bibfield  {journal} {\bibinfo  {journal} {J. Comput. Phys.}\ }\textbf {\bibinfo {volume} {179\normalfont{(2)}}},\ \bibinfo {pages} {495} (\bibinfo {year} {2002})}\BibitemShut {NoStop}%
\bibitem [{\citenamefont {Valentini}\ \emph {et~al.}(2011)\citenamefont {Valentini}, \citenamefont {Califano}, \citenamefont {Perrone}, \citenamefont {Pegoraro},\ and\ \citenamefont {Veltri}}]{Vl1}%
  \BibitemOpen
  \bibfield  {author} {\bibinfo {author} {\bibfnamefont {F.}~\bibnamefont {Valentini}}, \bibinfo {author} {\bibfnamefont {F.}~\bibnamefont {Califano}}, \bibinfo {author} {\bibfnamefont {D.}~\bibnamefont {Perrone}}, \bibinfo {author} {\bibfnamefont {F.}~\bibnamefont {Pegoraro}}, \ and\ \bibinfo {author} {\bibfnamefont {P.}~\bibnamefont {Veltri}},\ }\href@noop {} {\bibfield  {journal} {\bibinfo  {journal} {Phys. Rev. Lett.}\ }\textbf {\bibinfo {volume} {106\normalfont{(16)}}},\ \bibinfo {pages} {165002} (\bibinfo {year} {2011})}\BibitemShut {NoStop}%
\bibitem [{\citenamefont {Valentini}\ \emph {et~al.}(2012)\citenamefont {Valentini}, \citenamefont {Perrone}, \citenamefont {Califano}, \citenamefont {Pegoraro}, \citenamefont {Veltri}, \citenamefont {Morrison},\ and\ \citenamefont {O'Neil}}]{Vl2}%
  \BibitemOpen
  \bibfield  {author} {\bibinfo {author} {\bibfnamefont {F.}~\bibnamefont {Valentini}}, \bibinfo {author} {\bibfnamefont {D.}~\bibnamefont {Perrone}}, \bibinfo {author} {\bibfnamefont {F.}~\bibnamefont {Califano}}, \bibinfo {author} {\bibfnamefont {F.}~\bibnamefont {Pegoraro}}, \bibinfo {author} {\bibfnamefont {P.}~\bibnamefont {Veltri}}, \bibinfo {author} {\bibfnamefont {P.~J.}\ \bibnamefont {Morrison}}, \ and\ \bibinfo {author} {\bibfnamefont {T.~M.}\ \bibnamefont {O'Neil}},\ }\href@noop {} {\bibfield  {journal} {\bibinfo  {journal} {Phys. Plasmas}\ }\textbf {\bibinfo {volume} {19\normalfont{(9)}}},\ \bibinfo {pages} {092103} (\bibinfo {year} {2012})}\BibitemShut {NoStop}%
\bibitem [{\citenamefont {Valentini}\ \emph {et~al.}(2013)\citenamefont {Valentini}, \citenamefont {Perrone}, \citenamefont {Califano}, \citenamefont {Pegoraro}, \citenamefont {Veltri}, \citenamefont {Morrison},\ and\ \citenamefont {O'Neil}}]{Vl3}%
  \BibitemOpen
  \bibfield  {author} {\bibinfo {author} {\bibfnamefont {F.}~\bibnamefont {Valentini}}, \bibinfo {author} {\bibfnamefont {D.}~\bibnamefont {Perrone}}, \bibinfo {author} {\bibfnamefont {F.}~\bibnamefont {Califano}}, \bibinfo {author} {\bibfnamefont {F.}~\bibnamefont {Pegoraro}}, \bibinfo {author} {\bibfnamefont {P.}~\bibnamefont {Veltri}}, \bibinfo {author} {\bibfnamefont {P.~J.}\ \bibnamefont {Morrison}}, \ and\ \bibinfo {author} {\bibfnamefont {T.~M.}\ \bibnamefont {O'Neil}},\ }\href@noop {} {\bibfield  {journal} {\bibinfo  {journal} {Phys. Plasmas}\ }\textbf {\bibinfo {volume} {20\normalfont{(3)}}},\ \bibinfo {pages} {034702} (\bibinfo {year} {2013})}\BibitemShut {NoStop}%
\bibitem [{\citenamefont {Perrone}\ \emph {et~al.}(2013)\citenamefont {Perrone}, \citenamefont {Valentini}, \citenamefont {Servidio}, \citenamefont {Dalena},\ and\ \citenamefont {Veltri}}]{Vl4}%
  \BibitemOpen
  \bibfield  {author} {\bibinfo {author} {\bibfnamefont {D.}~\bibnamefont {Perrone}}, \bibinfo {author} {\bibfnamefont {F.}~\bibnamefont {Valentini}}, \bibinfo {author} {\bibfnamefont {S.}~\bibnamefont {Servidio}}, \bibinfo {author} {\bibfnamefont {S.}~\bibnamefont {Dalena}}, \ and\ \bibinfo {author} {\bibfnamefont {P.}~\bibnamefont {Veltri}},\ }\href@noop {} {\bibfield  {journal} {\bibinfo  {journal} {Astrophys. J.}\ }\textbf {\bibinfo {volume} {762\normalfont{(2)}}},\ \bibinfo {pages} {99} (\bibinfo {year} {2013})}\BibitemShut {NoStop}%
\bibitem [{\citenamefont {Cheng}\ and\ \citenamefont {Knorr}(1976)}]{ChengKnorr}%
  \BibitemOpen
  \bibfield  {author} {\bibinfo {author} {\bibfnamefont {C.~Z.}\ \bibnamefont {Cheng}}\ and\ \bibinfo {author} {\bibfnamefont {G.}~\bibnamefont {Knorr}},\ }\href@noop {} {\bibfield  {journal} {\bibinfo  {journal} {J. Comput. Phys.}\ }\textbf {\bibinfo {volume} {22\normalfont{(3)}}},\ \bibinfo {pages} {330} (\bibinfo {year} {1976})}\BibitemShut {NoStop}%
\bibitem [{\citenamefont {Valentini}\ \emph {et~al.}(2005{\natexlab{b}})\citenamefont {Valentini}, \citenamefont {Veltri},\ and\ \citenamefont {Mangeney}}]{split1}%
  \BibitemOpen
  \bibfield  {author} {\bibinfo {author} {\bibfnamefont {F.}~\bibnamefont {Valentini}}, \bibinfo {author} {\bibfnamefont {P.}~\bibnamefont {Veltri}}, \ and\ \bibinfo {author} {\bibfnamefont {A.}~\bibnamefont {Mangeney}},\ }\href@noop {} {\bibfield  {journal} {\bibinfo  {journal} {J. Comput. Phys.}\ }\textbf {\bibinfo {volume} {210\normalfont{(2)}}},\ \bibinfo {pages} {730} (\bibinfo {year} {2005}{\natexlab{b}})}\BibitemShut {NoStop}%
\bibitem [{\citenamefont {Valentini}\ \emph {et~al.}(2007)\citenamefont {Valentini}, \citenamefont {Travnicek}, \citenamefont {Califano}, \citenamefont {Hellinger},\ and\ \citenamefont {Mangeney}}]{split2}%
  \BibitemOpen
  \bibfield  {author} {\bibinfo {author} {\bibfnamefont {F.}~\bibnamefont {Valentini}}, \bibinfo {author} {\bibfnamefont {P.}~\bibnamefont {Travnicek}}, \bibinfo {author} {\bibfnamefont {F.}~\bibnamefont {Califano}}, \bibinfo {author} {\bibfnamefont {P.}~\bibnamefont {Hellinger}}, \ and\ \bibinfo {author} {\bibfnamefont {A.}~\bibnamefont {Mangeney}},\ }\href@noop {} {\bibfield  {journal} {\bibinfo  {journal} {J. Comput. Phys.}\ }\textbf {\bibinfo {volume} {225\normalfont{(1)}}},\ \bibinfo {pages} {753} (\bibinfo {year} {2007})}\BibitemShut {NoStop}%
\bibitem [{\citenamefont {Filbet}\ and\ \citenamefont {Pareschi}(2002)}]{VLnh}%
  \BibitemOpen
  \bibfield  {author} {\bibinfo {author} {\bibfnamefont {F.}~\bibnamefont {Filbet}}\ and\ \bibinfo {author} {\bibfnamefont {L.}~\bibnamefont {Pareschi}},\ }\href@noop {} {\bibfield  {journal} {\bibinfo  {journal} {J. Comput. Phys.}\ }\textbf {\bibinfo {volume} {179\normalfont{(1)}}},\ \bibinfo {pages} {1} (\bibinfo {year} {2002})}\BibitemShut {NoStop}%
\bibitem [{\citenamefont {Peyret}\ and\ \citenamefont {Taylor}(1983)}]{CFL}%
  \BibitemOpen
  \bibfield  {author} {\bibinfo {author} {\bibfnamefont {R.}~\bibnamefont {Peyret}}\ and\ \bibinfo {author} {\bibfnamefont {T.~D.}\ \bibnamefont {Taylor}},\ }\href@noop {} {\emph {\bibinfo {title} {Computational methods for fluid flow}}}\ (\bibinfo  {publisher} {Springer},\ \bibinfo {address} {New York},\ \bibinfo {year} {1983})\BibitemShut {NoStop}%
\bibitem [{\citenamefont {Press}\ \emph {et~al.}(1992)\citenamefont {Press}, \citenamefont {Teukolsky}, \citenamefont {Vetterling},\ and\ \citenamefont {Flannery}}]{Gauss}%
  \BibitemOpen
  \bibfield  {author} {\bibinfo {author} {\bibfnamefont {W.~H.}\ \bibnamefont {Press}}, \bibinfo {author} {\bibfnamefont {A.~W.}\ \bibnamefont {Teukolsky}}, \bibinfo {author} {\bibfnamefont {W.~T.}\ \bibnamefont {Vetterling}}, \ and\ \bibinfo {author} {\bibfnamefont {B.~P.}\ \bibnamefont {Flannery}},\ }\href@noop {} {\emph {\bibinfo {title} {Numerical recipes in C. The art of scientific computing}}},\ \bibinfo {edition} {2nd}\ ed.\ (\bibinfo  {publisher} {Cambridge University Press},\ \bibinfo {address} {Cambridge},\ \bibinfo {year} {1992})\ pp.\ \bibinfo {pages} {147--150}\BibitemShut {NoStop}%
\bibitem [{\citenamefont {Pezzi}\ \emph {et~al.}(2019{\natexlab{b}})\citenamefont {Pezzi}, \citenamefont {Perrone}, \citenamefont {Servidio}, \citenamefont {Valentini}, \citenamefont {Sorriso-Valvo},\ and\ \citenamefont {Veltri}}]{Pezziquad}%
  \BibitemOpen
  \bibfield  {author} {\bibinfo {author} {\bibfnamefont {O.}~\bibnamefont {Pezzi}}, \bibinfo {author} {\bibfnamefont {D.}~\bibnamefont {Perrone}}, \bibinfo {author} {\bibfnamefont {S.}~\bibnamefont {Servidio}}, \bibinfo {author} {\bibfnamefont {F.}~\bibnamefont {Valentini}}, \bibinfo {author} {\bibfnamefont {L.}~\bibnamefont {Sorriso-Valvo}}, \ and\ \bibinfo {author} {\bibfnamefont {P.}~\bibnamefont {Veltri}},\ }\href@noop {} {\bibfield  {journal} {\bibinfo  {journal} {Astrophys. J.}\ }\textbf {\bibinfo {volume} {887\normalfont{(2)}}},\ \bibinfo {pages} {208} (\bibinfo {year} {2019}{\natexlab{b}})}\BibitemShut {NoStop}%
\bibitem [{\citenamefont {Golub}\ and\ \citenamefont {Welsch}(1969)}]{quad}%
  \BibitemOpen
  \bibfield  {author} {\bibinfo {author} {\bibfnamefont {G.~H.}\ \bibnamefont {Golub}}\ and\ \bibinfo {author} {\bibfnamefont {J.~H.}\ \bibnamefont {Welsch}},\ }\href@noop {} {\bibfield  {journal} {\bibinfo  {journal} {Math. Comp.}\ }\textbf {\bibinfo {volume} {23\normalfont{(106)}}},\ \bibinfo {pages} {221} (\bibinfo {year} {1969})}\BibitemShut {NoStop}%
\bibitem [{\citenamefont {Cassak}\ \emph {et~al.}(2023)\citenamefont {Cassak}, \citenamefont {Barbhuiya}, \citenamefont {Liang},\ and\ \citenamefont {Argall}}]{Cassak}%
  \BibitemOpen
  \bibfield  {author} {\bibinfo {author} {\bibfnamefont {P.~A.}\ \bibnamefont {Cassak}}, \bibinfo {author} {\bibfnamefont {M.~H.}\ \bibnamefont {Barbhuiya}}, \bibinfo {author} {\bibfnamefont {H.}~\bibnamefont {Liang}}, \ and\ \bibinfo {author} {\bibfnamefont {M.~R.}\ \bibnamefont {Argall}},\ }\href@noop {} {\bibfield  {journal} {\bibinfo  {journal} {Phys. Rev. Lett.}\ }\textbf {\bibinfo {volume} {130\normalfont{(8)}}},\ \bibinfo {pages} {085201} (\bibinfo {year} {2023})}\BibitemShut {NoStop}%
\bibitem [{\citenamefont {Krall}\ and\ \citenamefont {Trivelpiece}(1973{\natexlab{d}})}]{Langmuir}%
  \BibitemOpen
  \bibfield  {author} {\bibinfo {author} {\bibfnamefont {N.~A.}\ \bibnamefont {Krall}}\ and\ \bibinfo {author} {\bibfnamefont {A.~W.}\ \bibnamefont {Trivelpiece}},\ }\href@noop {} {\emph {\bibinfo {title} {Principles of plasma physics}}}\ (\bibinfo  {publisher} {McGraw-Hill},\ \bibinfo {address} {New York},\ \bibinfo {year} {1973})\ pp.\ \bibinfo {pages} {383--395}\BibitemShut {NoStop}%
\bibitem [{\citenamefont {Anderson}\ \emph {et~al.}(2001)\citenamefont {Anderson}, \citenamefont {Fedele},\ and\ \citenamefont {Lisak}}]{tutorial}%
  \BibitemOpen
  \bibfield  {author} {\bibinfo {author} {\bibfnamefont {D.}~\bibnamefont {Anderson}}, \bibinfo {author} {\bibfnamefont {R.}~\bibnamefont {Fedele}}, \ and\ \bibinfo {author} {\bibfnamefont {M.}~\bibnamefont {Lisak}},\ }\href@noop {} {\bibfield  {journal} {\bibinfo  {journal} {Am. J. Phys.}\ }\textbf {\bibinfo {volume} {69\normalfont{(12)}}},\ \bibinfo {pages} {1262} (\bibinfo {year} {2001})}\BibitemShut {NoStop}%
\bibitem [{\citenamefont {Dawson}(1960)}]{TwoBeam}%
  \BibitemOpen
  \bibfield  {author} {\bibinfo {author} {\bibfnamefont {J.~M.}\ \bibnamefont {Dawson}},\ }\href@noop {} {\bibfield  {journal} {\bibinfo  {journal} {Phys. Rev.}\ }\textbf {\bibinfo {volume} {118\normalfont{(2)}}},\ \bibinfo {pages} {381} (\bibinfo {year} {1960})}\BibitemShut {NoStop}%
\bibitem [{\citenamefont {DePackh}(1962)}]{vortex1}%
  \BibitemOpen
  \bibfield  {author} {\bibinfo {author} {\bibfnamefont {D.~C.}\ \bibnamefont {DePackh}},\ }\href@noop {} {\bibfield  {journal} {\bibinfo  {journal} {J. Electron. Control}\ }\textbf {\bibinfo {volume} {13\normalfont{(5)}}},\ \bibinfo {pages} {417} (\bibinfo {year} {1962})}\BibitemShut {NoStop}%
\bibitem [{\citenamefont {Dory}(1964)}]{vortex2}%
  \BibitemOpen
  \bibfield  {author} {\bibinfo {author} {\bibfnamefont {R.~A.}\ \bibnamefont {Dory}},\ }\href@noop {} {\bibfield  {journal} {\bibinfo  {journal} {J. Nucl. Energy, Part C}\ }\textbf {\bibinfo {volume} {6\normalfont{(5)}}},\ \bibinfo {pages} {511} (\bibinfo {year} {1964})}\BibitemShut {NoStop}%
\bibitem [{\citenamefont {Berk}\ \emph {et~al.}(1970)\citenamefont {Berk}, \citenamefont {Nielsen},\ and\ \citenamefont {Roberts}}]{vortex3}%
  \BibitemOpen
  \bibfield  {author} {\bibinfo {author} {\bibfnamefont {H.~L.}\ \bibnamefont {Berk}}, \bibinfo {author} {\bibfnamefont {C.~E.}\ \bibnamefont {Nielsen}}, \ and\ \bibinfo {author} {\bibfnamefont {K.~V.}\ \bibnamefont {Roberts}},\ }\href@noop {} {\bibfield  {journal} {\bibinfo  {journal} {Phys. Fluids}\ }\textbf {\bibinfo {volume} {13\normalfont{(4)}}},\ \bibinfo {pages} {980} (\bibinfo {year} {1970})}\BibitemShut {NoStop}%
\bibitem [{\citenamefont {Ghizzo}\ \emph {et~al.}(1988)\citenamefont {Ghizzo}, \citenamefont {Izrar}, \citenamefont {Bertrand}, \citenamefont {Fijalkow}, \citenamefont {Feix},\ and\ \citenamefont {Shoucri}}]{vortex4}%
  \BibitemOpen
  \bibfield  {author} {\bibinfo {author} {\bibfnamefont {A.}~\bibnamefont {Ghizzo}}, \bibinfo {author} {\bibfnamefont {B.}~\bibnamefont {Izrar}}, \bibinfo {author} {\bibfnamefont {P.}~\bibnamefont {Bertrand}}, \bibinfo {author} {\bibfnamefont {E.}~\bibnamefont {Fijalkow}}, \bibinfo {author} {\bibfnamefont {M.~R.}\ \bibnamefont {Feix}}, \ and\ \bibinfo {author} {\bibfnamefont {M.}~\bibnamefont {Shoucri}},\ }\href@noop {} {\bibfield  {journal} {\bibinfo  {journal} {Phys. Fluids}\ }\textbf {\bibinfo {volume} {31\normalfont{(1)}}},\ \bibinfo {pages} {72} (\bibinfo {year} {1988})}\BibitemShut {NoStop}%
\bibitem [{\citenamefont {Manfredi}\ and\ \citenamefont {Bertrand}(2000)}]{vortex5}%
  \BibitemOpen
  \bibfield  {author} {\bibinfo {author} {\bibfnamefont {G.}~\bibnamefont {Manfredi}}\ and\ \bibinfo {author} {\bibfnamefont {P.}~\bibnamefont {Bertrand}},\ }\href@noop {} {\bibfield  {journal} {\bibinfo  {journal} {Phys. Plasmas}\ }\textbf {\bibinfo {volume} {7\normalfont{(6)}}},\ \bibinfo {pages} {2425} (\bibinfo {year} {2000})}\BibitemShut {NoStop}%
\bibitem [{\citenamefont {Valentini}\ \emph {et~al.}(2006)\citenamefont {Valentini}, \citenamefont {O'Neil},\ and\ \citenamefont {Dubin}}]{vortex6}%
  \BibitemOpen
  \bibfield  {author} {\bibinfo {author} {\bibfnamefont {F.}~\bibnamefont {Valentini}}, \bibinfo {author} {\bibfnamefont {T.~M.}\ \bibnamefont {O'Neil}}, \ and\ \bibinfo {author} {\bibfnamefont {D.~H.}\ \bibnamefont {Dubin}},\ }\href@noop {} {\bibfield  {journal} {\bibinfo  {journal} {Phys. Plasmas}\ }\textbf {\bibinfo {volume} {13\normalfont{(5)}}},\ \bibinfo {pages} {052303} (\bibinfo {year} {2006})}\BibitemShut {NoStop}%
\bibitem [{\citenamefont {Tavassoli}\ \emph {et~al.}(2023)\citenamefont {Tavassoli}, \citenamefont {Papahn~Zadeh}, \citenamefont {Smolyakov}, \citenamefont {Shoucri},\ and\ \citenamefont {Spiteri}}]{casc1}%
  \BibitemOpen
  \bibfield  {author} {\bibinfo {author} {\bibfnamefont {A.}~\bibnamefont {Tavassoli}}, \bibinfo {author} {\bibfnamefont {M.}~\bibnamefont {Papahn~Zadeh}}, \bibinfo {author} {\bibfnamefont {A.}~\bibnamefont {Smolyakov}}, \bibinfo {author} {\bibfnamefont {M.}~\bibnamefont {Shoucri}}, \ and\ \bibinfo {author} {\bibfnamefont {R.~J.}\ \bibnamefont {Spiteri}},\ }\href@noop {} {\bibfield  {journal} {\bibinfo  {journal} {Phys. Plasmas}\ }\textbf {\bibinfo {volume} {30\normalfont{(3)}}},\ \bibinfo {pages} {033905} (\bibinfo {year} {2023})}\BibitemShut {NoStop}%
\bibitem [{\citenamefont {Brown}\ and\ \citenamefont {Jorns}(2023)}]{casc2}%
  \BibitemOpen
  \bibfield  {author} {\bibinfo {author} {\bibfnamefont {Z.~A.}\ \bibnamefont {Brown}}\ and\ \bibinfo {author} {\bibfnamefont {B.~A.}\ \bibnamefont {Jorns}},\ }\href@noop {} {\bibfield  {journal} {\bibinfo  {journal} {Phys. Rev. Lett.}\ }\textbf {\bibinfo {volume} {130\normalfont{(11)}}},\ \bibinfo {pages} {115101} (\bibinfo {year} {2023})}\BibitemShut {NoStop}%
\bibitem [{\citenamefont {Kolmogorov}(1941)}]{Kol}%
  \BibitemOpen
  \bibfield  {author} {\bibinfo {author} {\bibfnamefont {A.~N.}\ \bibnamefont {Kolmogorov}},\ }\href@noop {} {\bibfield  {journal} {\bibinfo  {journal} {Dokl. Akad. Nauk SSSR}\ }\textbf {\bibinfo {volume} {30}},\ \bibinfo {pages} {301} (\bibinfo {year} {1941})}\BibitemShut {NoStop}%
\bibitem [{\citenamefont {Pezzi}\ \emph {et~al.}(2016{\natexlab{b}})\citenamefont {Pezzi}, \citenamefont {Valentini},\ and\ \citenamefont {Veltri}}]{pezzicoll}%
  \BibitemOpen
  \bibfield  {author} {\bibinfo {author} {\bibfnamefont {O.}~\bibnamefont {Pezzi}}, \bibinfo {author} {\bibfnamefont {F.}~\bibnamefont {Valentini}}, \ and\ \bibinfo {author} {\bibfnamefont {P.}~\bibnamefont {Veltri}},\ }\href@noop {} {\bibfield  {journal} {\bibinfo  {journal} {Phys. Rev. Lett.}\ }\textbf {\bibinfo {volume} {116\normalfont{(14)}}},\ \bibinfo {pages} {145001} (\bibinfo {year} {2016}{\natexlab{b}})}\BibitemShut {NoStop}%
\bibitem [{\citenamefont {Abramowitz}\ and\ \citenamefont {Stegun}(1972{\natexlab{a}})}]{AbrSte1}%
  \BibitemOpen
  \bibfield  {author} {\bibinfo {author} {\bibfnamefont {M.}~\bibnamefont {Abramowitz}}\ and\ \bibinfo {author} {\bibfnamefont {I.~A.}\ \bibnamefont {Stegun}},\ }\href@noop {} {\emph {\bibinfo {title} {Handbook of mathematical functions with formulas, graphs, and mathematical tables}}},\ \bibinfo {edition} {10th}\ ed.\ (\bibinfo  {publisher} {United States Department of Commerce},\ \bibinfo {address} {Washington, DC},\ \bibinfo {year} {1972})\ \bibinfo {note} {eqs.~(13.5.16) and (13.6.38)}\BibitemShut {NoStop}%
\bibitem [{\citenamefont {Abramowitz}\ and\ \citenamefont {Stegun}(1972{\natexlab{b}})}]{AbrSte2}%
  \BibitemOpen
  \bibfield  {author} {\bibinfo {author} {\bibfnamefont {M.}~\bibnamefont {Abramowitz}}\ and\ \bibinfo {author} {\bibfnamefont {I.~A.}\ \bibnamefont {Stegun}},\ }\href@noop {} {\emph {\bibinfo {title} {Handbook of mathematical functions with formulas, graphs, and mathematical tables}}},\ \bibinfo {edition} {10th}\ ed.\ (\bibinfo  {publisher} {United States Department of Commerce},\ \bibinfo {address} {Washington, DC},\ \bibinfo {year} {1972})\ \bibinfo {note} {eqs.~(6.1.37)-(6.1.38)}\BibitemShut {NoStop}%
\bibitem [{\citenamefont {Brush}(1966)}]{Boltzmann}%
  \BibitemOpen
  \bibfield  {author} {\bibinfo {author} {\bibfnamefont {S.~G.}\ \bibnamefont {Brush}},\ }\href@noop {} {\emph {\bibinfo {title} {Kinetic theory. Volume 2, Irreversible processes}}}\ (\bibinfo  {publisher} {Pergamon},\ \bibinfo {address} {Oxford},\ \bibinfo {year} {1966})\BibitemShut {NoStop}%
\end{thebibliography}%

\end{document}